\documentclass[twocolumn]{aastex631}

\usepackage{amsmath}
\usepackage{breqn}
\usepackage{bm}
\usepackage{dsfont}
\usepackage{hyperref}
\usepackage{makecell}

\begin{document}

\title{Stochastic analysis of ultra-high energy cosmic ray interactions}

\correspondingauthor{Leonel Morejon}
\email{leonel.morejon@uni-wuppertal.de}

\author[0000-0003-1494-2624]{Leonel Morejon}
\affiliation{Bergische Universit\"{a}t Wuppertal \\
Gaußstraße 20,
42119 Wuppertal, Germany}

\author[0000-0002-2805-0195]{Karl-Heinz Kampert}
\affiliation{Bergische Universit\"{a}t Wuppertal \\
Gaußstraße 20,
42119 Wuppertal, Germany}

\begin{abstract}

Photonuclear interactions between ultra-high-energy cosmic ray (UHECR) nuclei and surrounding photon fields are key to understanding the connection between the compositions observed at Earth and those emitted from the sources. These interactions can completely disintegrate a nucleus of iron over trajectory lengths of a few and up to hundreds of megaparsecs, depending on the energy of the UHECR. The stochastic nature of these interactions means that it is not possible to describe them deterministically for a single cosmic ray, and an exact formulation of the probability distributions is not yet available. Current approaches describe these interactions using either Monte Carlo simulations or solving ordinary differential equations that neglect stochasticity. Because of the limitations of these approaches, only partial capture of the process is achieved.  This paper presents an analytic probabilistic description of UHECR interactions and the resulting nuclear cascades, establishing their connection to Markov jump processes. The fundamental properties of these cascades are presented, as is the computation of the usual quantities of interest, such as the horizon, spectrum, and composition. The benefits of this description are outlined using astrophysical examples related to extragalactic propagation and UHECR sources.

\end{abstract}


\section{Introduction} 
\label{sec:intro}

Experimental observations of cosmic rays alone are insufficient to answer the fundamental questions about their origins. A precise understanding of the magnetic deflections and interactions that affect their production and propagation is essential for reconstructing their past history. In the case of ultrahigh-energy cosmic rays (UHECRs), it is now understood that their sources must be extragalactic~\citep{Aab_2015,Abdul_Halim_2024b}. The plausibility of hypothetical sources is assessed by using knowledge of interactions and magnetic deflections to produce synthetic quantities that can be compared with the main observables, such as the energy spectrum and fluctuations in the depth of shower maximum \citep{Abdul_Halim_2023b} as well as, more recently, including arrival directions \citep{Abdul_Halim_2024}. The present work focuses on the interactions of UHECRs, with some mention of the effects of turbulent magnetic fields. The effect of Galactic magnetic fields~\citep{Unger_2024a,Korochkin:2024yit} will not be addressed here.

During acceleration and diffusion within the sources, as well as during propagation, UHECRs interact with surrounding photon fields\,\footnote{Hadronic interactions are less important for UHECRs but may also occur. A subsequent publication will discuss the stochastic analytic method presented here for hadronic interactions.}, such as the cosmic microwave background (CMB), the cosmic infrared background (IRB), or nonthermal spectra in the source. These interactions result in the loss of energy and photodisintegration of the UHECR. Some interactions do not change the nuclear species of the UHECR and are well characterized as deterministic (often referred to as continuous energy losses, CEL) if fluctuations of the inelasticity and the interaction length are negligible. Some examples include Bethe-Heitler pair production and synchrotron losses. Conversely, stochastic interactions often result in the transformation or loss of the interacting particle (referred to as stochastic losses, SL) with discrete changes in interaction lengths and multiple possible outcomes for the resulting species. Examples of such interactions include the photodisintegration of cosmic ray nuclei, where the number of nucleons lost is not deterministic, as well as photomeson production, where multiple meson-producing channels are available (depending on the energy), each with a distribution of inelasticity and number of secondaries. This process also leads to nuclear fragments~\citep{Morejon2019}.

Although the fundamental differences between SL and CEL were already recognized in early works \citep{Puget1976,1993PThPh..89..833Y}, a CEL approach has been widely employed (e.g.~\cite{Hill1985,1988A&A...199....1B}) to describe the evolution of the cosmic ray spectrum. However, improvements in the experimental precision and indications of a heavier composition have increased the need for more sophisticated methods that account for the probabilistic nature of SL efficiently. Today, approaches to computing the interactions of UHECR nuclei can be grouped into two types. The first type uses the continuous-limit approximation \citep{PhysRevD.77.103007,ALOISIO201373,ALOISIO201394,Boncioli:2016lkt,Biehl2018CosmicCascade,Heinze:2019jou,Lia_2024}, where the energy densities of different nuclear species are computed by solving a coupled system of differential equations in which SL and CEL are treated as continuous losses. The second type uses Monte Carlo methods \citep{1989BAAS...21..780H,Epele_1998,10.1093/mnras/stv893,Batista2016,Aloisio_2017}, which simulate the underlying stochasticity by sampling each interaction and tracking the products individually. The former approach has the advantage of faster computation and analytic solutions have even been put forward by limiting the number of disintegration channels~\citep{PhysRevD.77.103007,PhysRevD.82.123005,Ahlers2013,ALOISIO201373,ALOISIO201394}.
The latter is considered to be more theoretically correct because it best reflects the nature of the interactions and allows for the estimation of stochastic effects. However, there are intrinsic limitations to the method, such as the computational expense involved, depending on assumptions, and problems of convergence~\citep[e.g.,][]{asmussen2007stochastic}. More importantly, Monte Carlo simulations provide limited theoretical insight because the impact of input uncertainties (e.g.,~nuclear cross-sections and photon field models) cannot be easily determined without an exhaustive and computationally demanding parameter space scan. In contrast, an analytic framework can facilitate the study of correlations between inputs (in some cases, explicitly) and the computation is considerably more efficient. It can also achieve arbitrary precision at modest computational effort. Conversely, Monte Carlo methods often waste computational resources on uninteresting events as they are blind to the underlying probability space (see Appendix~\ref{sec:efficiency}). Currently, there is no formal theoretical framework that can describe the stochasticity of the UHECR interactions analytically. 

This paper presents an analytic theoretical framework that addresses the interactions of UHECRs with photon fields prevalent in extragalactic propagation and within sources. The resulting closed-form expressions describe the probability distributions as a function of target thickness for an arbitrary initial condition. This approach can easily be extended to include nuclear masses beyond iron, enabling the independent study of the effects of uncertainties in inputs such as nuclear cross-sections and photon fields.

\section{Stochastic description}
\label{sec:stochastic_losses}

The continuous-limit temporal evolution of the energy densities of UHECR nuclei interacting with photon fields is described by a coupled system of ordinary differential equations,
\begin{equation}
    \frac{\partial}{\partial t} n_i (E_i) = \frac{\partial}{\partial E_i} (b n_i) + q_i^{\mathrm{ext}}(E_i)  + \sum_j \lambda_{j\to i}(E_j) n_j(E_j),
    \label{eq:standard_SODE}
\end{equation} 
where $n_i (E_i) \equiv  n_i (E_i, t)$ is the differential number density of nuclear species $i$ (noting that quantities are time dependencies although not written explicitly). The first term on the right-hand side includes all CEL processes, such as synchrotron and escape losses in the case of source scenarios, and pair production and adiabatic losses in the case of extragalactic propagation. The term $q_i^{\mathrm{ext}}$ describes the injection of particles with energy, $E_i$, which could represent the injections through acceleration mechanisms within the sources, or emissions from multiple sources in the case of extragalactic propagation. 
The terms $\lambda_{j\to i}(E_j)$ denote the interaction rates for all SL processes incurred by species $j$, leading to the production of species $i$, with photons of energy $\epsilon$ and number density $n(\epsilon)$. This includes the term  $j=i$, which has a different form $-\lambda_i n_i (E_i)$ with $\lambda_i(E_i)$, the total interaction rate. The total cross-section for species $i$ as a function of photon energy, $\varepsilon$, (in the center-of-mass rest frame), $\sigma_i(\varepsilon) = \sum_k \sigma_{i \to k}(\varepsilon)$, includes all possible products, $k$, and is given by the sum of the interaction rates $\lambda_i(\gamma) = \sum_k \lambda_{i \to k}(\gamma)$, which are computed as
\begin{equation}
    \lambda_{i\to k}(\gamma) = \frac{1}{2\gamma^2}\int_0^{\infty} \frac{n(\epsilon)}{\epsilon^2} d\epsilon \int_0^{2\epsilon \gamma} \varepsilon \sigma_{i \to k} (\varepsilon) d\varepsilon \, ,
    \label{eq:interaction_rate}
\end{equation}
with $\gamma$ representing the Lorenz factor. The system described by Eq.\,\ref{eq:standard_SODE} may comprise between $\sim$50-200 nuclear species when including elements up to iron and an energy grid with enough resolution ($\sim$100 bins in logarithmic scale) to capture the details of the spectra. 

The approach presented here aims at describing nuclear cascades initiated by individual cosmic rays and because of the boost preserving property of SLs, this implies solving Eq.~\ref{eq:standard_SODE} for individual values of the Lorentz boost $\gamma \approx E_k/m_k$, so we can write
 \begin{equation}
    \frac{\partial}{\partial t} \tilde n_i (\gamma) = \frac{\partial}{\partial \gamma} (\tilde b \tilde n_i) + \tilde q_i^{\mathrm{ext}}(\gamma)  + \sum_j \lambda_{j \to i}(\gamma) \tilde n_j(\gamma) ,
    \label{eq:boost_SODE}
    \end{equation}
where the tilde reflects that the quantities are now differential in boost instead of energy. This linear system of ordinary differential equations can be written as a matrix differential equation,
\begin{equation}
    \frac{\partial}{\partial t}\boldsymbol{N} - \boldsymbol{N}\boldsymbol{\Lambda}  = \frac{\partial}{\partial \gamma} (\tilde b \boldsymbol{N}) + \boldsymbol{Q}^{\mathrm{ext}} ,
    \label{eq:matrix_SODE}
\end{equation}
where $\boldsymbol{N}$ is a row vector containing all densities $\{\tilde n_i(\gamma, t)\}$, $\boldsymbol{Q}^{\mathrm{ext}}$ is a row vector, with elements $\{ \tilde q_i^{\mathrm{ext}}(\gamma) \}$, and $\boldsymbol{\Lambda}$ is the interaction rate matrix, $\{\lambda_{ji}=\lambda_{j \to i}(\gamma)\}$ ($\{\lambda_{ii}=-\lambda_i(\gamma)\}$), which is a square matrix with zeros for elements $j$ with no production of element $i$. The numerical integration of Eq.~\ref{eq:standard_SODE} (or, equivalently, Eq.~\ref{eq:matrix_SODE}) yields the time evolution of the species densities $\boldsymbol{N}(t)$ requiring knowledge of the initial densities $\boldsymbol{N}(t=0)\equiv \boldsymbol{N}_0$ and injections $\boldsymbol{Q}^{\mathrm{ext}}$ (including possible time dependence). Notably, for certain functions $\boldsymbol{Q}^{\mathrm{ext}}$, the solution may have an analytic form when the term $\frac{\partial}{\partial \gamma} (\tilde b \boldsymbol{N})$ is negligible (no CEL) since this term is the only one coupling the equations corresponding to different values of $\gamma$.

Equation~\ref{eq:matrix_SODE} (as Eq~\ref{eq:standard_SODE}) reflects the mean behavior of individual cascades (continuous limit), but it does not describe the stochastic behavior of the interactions and the resulting fluctuations of the stochastic quantities. The accurate underlying process is as follows: an initial UHECR nucleus propagates along a path of random length (determined by the relevant magnetic field) until it decays or interacts with the surrounding photon field. Each interaction produces a random number of secondaries according to a given set of probabilities. The secondaries and the remnant species (the secondary with the largest mass) continue to propagate under the influence of magnetic fields and subsequently interact stochastically with further random products. This corresponds to a Markov jump process~\citep{Bladt2017}, where the transient states are the nuclear species with transition probabilities determined by the current state. The transitions (jumps) are exponentially distributed as a function of the path length (or time). In this probabilistic framework, the homogeneous form of Eq.~\ref{eq:matrix_SODE} (without CEL and no injections) is analogous to Kolmogorov's differential equation,
\begin{equation}
    \frac{d}{d t} \boldsymbol{P}^t = \boldsymbol{G} \boldsymbol{P}^t = \boldsymbol{P}^t\boldsymbol{G} ,
    \label{eq:KolmogorovEqs}
\end{equation}
where instead of the density vector, $\boldsymbol{N}$, the more appropriate $\boldsymbol{P}^t$ appears, which is a matrix where each row $i$ contains the probabilities $p^t_{ij}$ of transitioning from the $i$-th state to each possible state $j$. at a time $t$. The infinitesimal generator, $\boldsymbol{G}$, is equivalent to the interaction matrix, $\boldsymbol{\Lambda}$, when no "absorbing state" is considered and fulfills $\boldsymbol{G}\boldsymbol{1}=\boldsymbol{0}$, where $\boldsymbol{1}$ and $\boldsymbol{0}$ are column vectors of ones and zeros, respectively, of the same dimension as $\boldsymbol{G}$. An "absorbing state" in our context is a species or set of species we wish to observe; thus, it does not interact or decay further once it is reached. This is of relevance to obtain the distance distributions until observing sets of species of interest (distance until absorption). When considering absorption, 
\begin{equation}
    \boldsymbol{G}(\gamma) = 
    \begin{pmatrix}
        \boldsymbol{\Lambda} & -\boldsymbol{\Lambda} \boldsymbol{1} \\
        \boldsymbol{0} & 0 \\
    \end{pmatrix} ,
    \label{eq:generator}
\end{equation}
where $\boldsymbol{1}$ ($\boldsymbol{0}$) are a column (row) vector of ones (zeros) of the same dimension as $\boldsymbol{\Lambda}$. This reflects that $\boldsymbol{G}$ contains the absorbing state, in addition to all species included in $\boldsymbol{\Lambda}$, and, consequently, the rows in $\boldsymbol{\Lambda}$ do not add up to zero for species which can produce species in the absorbing state. For these species, the main diagonal terms in $\boldsymbol{\Lambda}$, \{$-\lambda_i$\} contain also jumps to the absorbing state; hence, the last column $-\boldsymbol{\Lambda} \boldsymbol{1}$ to ensure $\boldsymbol{G}$ is properly defined.

The connection between Eq.~\ref{eq:KolmogorovEqs} and the homogeneous form of Eq.~\ref{eq:matrix_SODE} is evident since the solution $\boldsymbol{P}^t = e^{\boldsymbol{G}t}$ for time-homogeneous conditions (length and time independence of $\boldsymbol{\Lambda}$) in the former, is equivalent to the solution of the latter $\boldsymbol{N}(t) = \boldsymbol{N}_0e^{\boldsymbol{\Lambda}t}$, and, as mentioned $e^{\boldsymbol{G}t} \equiv e^{\boldsymbol{\Lambda}t}$ in the absence of any absorbing states. However, it should be emphasized that the two equations are in fact describing different quantities and are not completely equivalent: Eq.~\ref{eq:KolmogorovEqs} describes the time evolution of stochastic quantities, such as the occupation probability for each state of the nuclear cascade. Meanwhile, Eq.~\ref{eq:matrix_SODE} describes the time evolution of the deterministic quantities (the number density distribution for each nuclear species). These two descriptions can be connected in the continuous limit, when stochastic effects are less important and the evolution of the system behaves similarly to a fluid flow between a network of containers \citep[Section~4.6]{Bladt2017}. On the other hand, the differences between approaches are more appreciable for inhomogeneous scenarios or noncontinuous injection. For example, these descriptions may be equivalent for a delta-like injection term, $\boldsymbol{Q}^{\mathrm{ext}}$, but for a time-varying injection, the stochastic treatment needs additional assumptions (see Section~\ref{subsec:sources}). Similarly, the inclusion of CEL leads to inhomogeneities; thus, $\boldsymbol{P}^t \neq e^{\boldsymbol{G}t}$ and the equivalence of the approaches is not as evident as in the homogeneous case.

Nevertheless, the stochastic description presented here is not limited to stationary and homogeneous cases, changes in time or in the propagation path of UHECRs can be treated as a type of time-inhomogeneity; for instance,~a global change in the normalization of the target photon field density (see Section~\ref{sec:continuous_losses}). This is also how CELs are incorporated within this framework and other effects, such as the influence of magnetic fields. Before addressing the more complex inhomogeneous scenarios, it is helpful to outline the fundamental properties of the homogeneous ones and establish benchmarks. To this end, we focus here on the distributions of distance until reaching the absorption state.

\subsection{Serial and regular cascades; the canonical form}
\label{subsec:regular_serial}

First, we consider the case with only one nuclear species for each mass and only one possible interaction channel at each state; alternatively, there could be multiple channels, but some $m$-th channel has the largest branching ratio $\lambda_i(\gamma) \approx \lambda_{i \to m}(\gamma)$. A typical case is the one-nucleon-loss assumption~\citep{PhysRevD.77.103007}, where the cascade of a nucleus with mass $A$ proceeds in a chain of nuclei with descending mass $\{A, A-1, A-2, ..., A-k+2, A-k+1 \}$, denoting the sequence of states visited over $k$ consecutive interactions. The interactions of the species are governed by the respective rates, making up the interaction vector $\boldsymbol{\lambda}_{A \to A-k}(\gamma) := \{\lambda_A(\gamma), \lambda_{A-1}(\gamma), \lambda_{A-2}(\gamma), ..., \lambda_{A-k+2}(\gamma), \lambda_{A-k+1}(\gamma) \}$, computed by substituting the relevant cross-section into Eq.~\ref{eq:interaction_rate}, and evaluated on the common boost, $\gamma$. 
The sequential nature of these cascades implies that the probability distribution of the propagation path lengths until $k$ disintegrations, $L_k$, is the convolution of $k$ exponential distributions. This is a hypoexponential distribution with parameter vector $\boldsymbol{\lambda}(\gamma)$. The expected value of this distribution has a straightforward physical meaning: $\mathbb{E}[L_k] = \sum_{i=A-k+1}^A 1/\lambda_i$, the sum of the mean interaction lengths of each species in the chain. This assumption was used in \cite{MorejonPhdThesis} to understand the behavior of more complex disintegration networks for nuclei of masses up to lead. Cascades of these type,  in which each interaction produces only one channel with one nuclear species at each stage are referred to as serial cascades (SeCs) herein.

\begin{figure}[t]
\includegraphics[width=.45\textwidth]{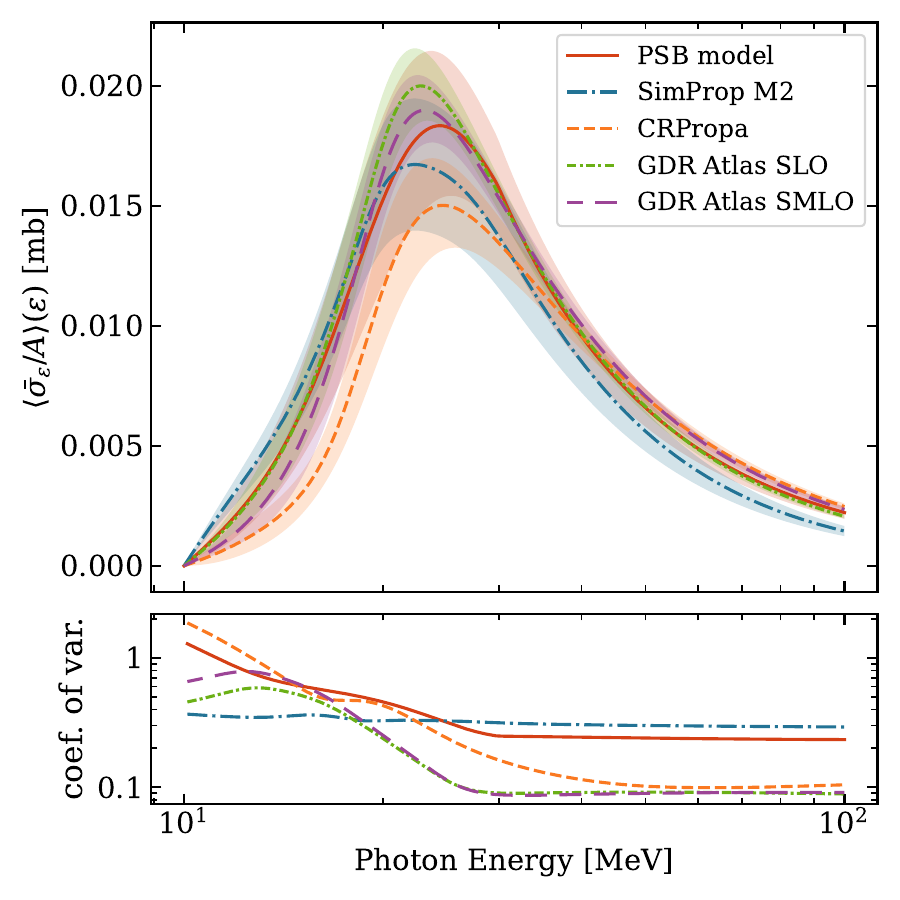}
\caption{Estimation of the deviation from regularity of photonuclear cross-sections. Top: Dependence of the energy-weighted photodisintegration cross-section divided by the nuclear mass as a function of photon energy. The lines show the average over all nuclear species in the respective model. The shaded bands represent the standard deviation at each energy and are centered on the mean. Bottom: The coefficient of variation (standard deviation divided by the mean) at each energy.}
\label{fig:rates_variations}
\end{figure}

For the canonical cascade, we assume that the photonuclear interaction rates are proportional to the mass number of the species, namely, $\lambda_A(\gamma) = A\lambda_1(\gamma)$, where $\lambda_1(\gamma)$ is the interaction rate per nucleon, which implies the relations $\lambda_{A_l} = \frac{A_l}{A_k} \lambda_{A_k}$ for any $k$ and $l$. This is motivated by the approximate proportionality of the photonuclear cross-section to the mass number, as reflected by the Thomas-Reiche-Kuhn sum rule ($\int \sigma(\varepsilon) d\varepsilon \propto \frac{ZN}{A}$) for giant dipole resonances (GDR) and the mass scaling of the cross-section in photomeson interactions~\citep{Morejon2019}. Cascades where the rates follow this type of proportionality with mass are called regular herein.
In general, photonuclear cross-sections deviate from this behavior from one species to another. However, these relations are a good approximation of the mean interaction rates and constitute a suitable benchmark for analyzing realistic distributions (see Fig.~\ref{fig:rates_variations}). 

Thus, we define the canonical form that we use as a benchmark, the regular serial cascade (RSeC): a SeC that obeys the regularity condition. The probability density of distances until $k$ nucleons are stripped from a nucleus of mass number $A$ is given by (see Appendix~\ref{sec:deriving_canonical}):
\begin{equation}
    f^{\mathrm{RS}}_{A \to A-k}(L) = \lambda_{A-k+1}e^{-\lambda_{A-k+1}L} \binom{A}{k-1} \left(1 - e^{-\lambda_1 L}\right)^{k-1} .
    \label{eq:kan_density}
\end{equation}

The interpretation of this expression is very intuitive: the distribution consists of $k$ independent events: the probability that any $k-1$ nucleons out of the initial $A$ interact within the trajectory length, $L$, (the term $ {A \choose k-1} \left(1 - e^{-\lambda_1 L}\right)^{k-1}$) and the probability density for the interaction of species with mass $A-k+1$ (the term $\lambda_{A-k+1}e^{-\lambda_{A-k+1}L} $), which is the last species that leads to the production of $A-k$. This interpretation becomes clearer in terms of the binomial distribution. Setting the interaction probability (success) for one nucleon to be equal to $\xi=1-e^{-\lambda_1L}$ the equation can be transformed to
\begin{equation}
    f^{\mathrm{RS}}_{A \to A-k}(\xi)=\frac{k}{\xi} \mathrm{B}(A, k, \xi)=\frac{A-k+1}{1-\xi} \mathrm{B}(A, k-1, \xi) ,
    \label{eq:kan_density_binomial}
\end{equation}
where the relation $\frac{d}{dL} = \lambda_1 (1 - \xi) \frac{d}{d\xi}$ is used. The binomial distribution, denoted by $\mathrm{B}(A, k, \xi)$, is the probability of obtaining $k$ disintegrations (successes) out of $A$ independent trials. This is a consequence of the regularity of the cascade: the constancy of the interaction rate per nucleon, $\lambda_1$, implies that nuclear effects are negligible and that the cascade is insensitive to the specific nuclei involved. The factors $\frac{k}{\xi}$, $\frac{A-k+1}{1-\xi}$ result from the change of differential variable in the density and the arbitrary choice of the ``success'' probability, $\xi$ or $1-\xi$.

Equation~\ref{eq:kan_density} is also equivalent to the beta distribution $\mathcal{B}(\alpha, \beta)$ with parameters $(\alpha=k, \beta=A-k+1)$ and defined expressions for the moments from which trivial relations for the RSeCs can be obtained (see Table~\ref{tab:moments}).
Given the previous expressions, the distribution for a specified initial composition, represented by the set of fractions $\{C_i\}=\{\eta_i, A_i\}$, where the fractions $\eta_i$ add up to 1, can be constructed as a linear combination of the distributions for each initial mass,
\begin{equation}
    f^{\mathrm{RS}}_{\{C_i\} \to A_f}(\xi) = \sum_i \eta_i f^{\mathrm{RS}}_{A_i \to A_f}(\xi) .
    \label{eq:canonical_mixture}
\end{equation}
\begin{table}[]
    \centering
    \begin{tabular}{c|c|c|c}
      C. type & Mean& Mode & Variance \\ \Xhline{1.0pt}
  & & &\\[-.7em]
  RSeC& $\lambda_1\ln \left( \frac{A+1}{A-k+1} \right)$& $\lambda_1\ln{\left( \frac{A-1 }{A-k}\right)}$& $-\lambda_1\ln \left( \frac{(A+1)^2(A+2)}{(A-k+1)k} \right)$\\ 
  & & &\\[-.7em]
  \hline
  & & &\\[-.7em]
  ISeC& $\sum_{i=A-k+1}^A 1/\lambda_i$& - & $\sum_{i=A-k+1}^A 1/\lambda_i^2$\\ 
  & & &\\[-.7em]
  \hline
  & & &\\[-.7em]
  CoC& $-\boldsymbol{\phi} \boldsymbol{\Lambda}^{-1}\boldsymbol{1}$ & - & $2\boldsymbol{\phi} \boldsymbol{\Lambda}^{-2}\boldsymbol{1} - (\boldsymbol{\phi} \boldsymbol{\Lambda}^{-1}\boldsymbol{1})^2$\\
\end{tabular}
    \caption{
    Characteristics of the distributions with a closed form.}
    \label{tab:moments}
\end{table}
\subsection{Irregular cascades and the nuclear decays}
\label{subsec:irregular_cascades}

The regularity condition assumes that nuclear cross-sections are unaffected by nuclear effects. In reality, changes in the number of protons and neutrons have a significant impact on the properties of the GDR, including the peak energy and the width. Consequently, the mass scaling of the interaction rates exhibits deviations from the regular values\footnote{\label{note1}This possibility also implies that the interaction matrix may be defective for a number of boosts, since different nuclei may have the same rate for some boost values. To avoid numerical problems, the corresponding boost values can be identified and excluded from the computation. The probability distributions for these values can then be determined by interpolating between valid adjacent boosts.}.

The deviations from regularity can be quantified independently of the target photon spectrum using the energy-weighted cross-section
\begin{equation}
    \bar \sigma_{\varepsilon}(\varepsilon) = \frac{2}{\varepsilon^2} \int_0^{\varepsilon} \varepsilon' \sigma (\varepsilon') d\varepsilon' \, ,
    \label{eq:energ_averaged_xsec}
\end{equation}
which forms part of Eq.~\ref{eq:interaction_rate} when rewritten in the form $\lambda(\gamma) = \int_0^{\infty} n(\epsilon) \bar\sigma_{\varepsilon}(2\gamma\epsilon)d\epsilon$. Figure~\ref{fig:rates_variations} represents the deviations from regularity for different cross-section datasets as the average over all nuclear species of the energy-weighted cross-section divided by the mass number. The shaded bands represent one standard deviation centered around the mean of the respective curves, and the bottom plot shows the coefficient of variation (ratio of width of the band to the line values). For a regular model the bands collapse to the mean line, since the standard deviation would be null. The models shown illustrate different existing choices for the set of nuclear species and the functional shape of their cross-sections: some contain only one species per mass number , as the PSB model~\citep{Puget1976} or the model available in SimProp~v2r4~\citep{Aloisio_2017} with command-line option \texttt{-M 2 < xsect_BreitWigner_TALYS-1.6.txt}, both with 56 species; meanwhile, others contain larger collections of species, such as the default model in CRPropa~3.2~\citep{Kampert:2012fi,2022JCAP...09..035A} with 184 species, or the much larger collection of cross-sections, the GDR Atlas~\citep{Kawano2020}, which has two different parameterizations for the GDR (SLO / SMLO) and covers 532 species up to nuclear mass 56 (larger masses are also available in the Atlas). The coefficient of variation is large for energies below the GDR and reduces after the peak for all models, typically to about 10\% or less for all except the serial models which remain above 30\%. The mean energy-weighted cross-section divided by the mass number is a fundamental quantity for a cross-section model, as it is connected to the mean interaction rate per nucleon by $\langle\lambda_1\rangle(\gamma) = \int_0^{\infty} n(\epsilon) \langle\bar\sigma_{\varepsilon}/A\rangle(2\gamma\epsilon)d\epsilon$ which is the analogous of $\lambda_1$ in irregular models.

Another cause of irregularity is spontaneous nuclear decay because in this framework the decay rate is part of the total interaction rate, 
\begin{equation}
    \lambda^{\mathrm{tot}}_{A_i \to A_j}(\gamma) = \lambda_{A_i \to A_j} (\gamma) + \gamma /c \tau ,
    \label{eq:inclusive_rates}
\end{equation}
which produces deviations from regularity for boost values and decay times, $\tau$, where the second term is comparable to the first. SeCs with rates that deviate from the regular relations are referred to as irregular serial cascades (ISeCs) herein.

In contrast to Eq.~\ref{eq:kan_density_binomial}, the probability density for ISeCs cannot be reduced to a dependence on the masses, because the mass scaling regularity does not apply. An ISeC is distributed according to  a hypoexponential distribution, while its density can be expressed as a linear combination of the exponential distributions with interaction vector $\boldsymbol{\lambda}_{A \to A-k}(\gamma)$ as long as they are all different\footnote{See footnote \ref{note1}.},
\begin{equation}
    f^{\mathrm{IS}}_{A \to A-k}(L) = \sum_{i=1}^{k} p_i(0) \lambda_{A-i} e^{(-\lambda_{A-i} L)} .
    \label{eq:isec_density_distinct}
\end{equation}
Here the coefficients $p_i(0)$ are the Lagrange interpolation polynomials evaluated at $\lambda=0$,
\begin{equation}
    p_j(\lambda) = \prod_{j=1,\, j \ne i}^{k} \frac{\lambda_{A-j} - \lambda}{\lambda_{A-j} - \lambda_{A-i}} .
\end{equation}
Equation~\ref{eq:isec_density_distinct} facilitates estimating the impact of irregularity on the density and it can be reduced to Eq.~\ref{eq:kan_density} when the regularity condition is imposed, as expected (see Appendix~\ref{sec:expected_distance}).

The more general expression, which is also applicable to cases where not all rates differ, is
\begin{equation}
    f^{\mathrm{IS}}_{A \to A-k}(L)= -\boldsymbol{\phi} e^{\boldsymbol{\Lambda} L } \boldsymbol{\Lambda} \boldsymbol{1} ,
    \label{eq:isec_density}
\end{equation}
where $\boldsymbol{\phi}$ is a row vector denoting the initial fractions. Therefore, the vector is all zeros except for a one in the first element, as in this case, there is only one starting species which corresponds to mass $A$. The interaction matrix $\boldsymbol{\Lambda}$ contains the negative interaction rates $\boldsymbol{\lambda}\equiv\{\lambda_A(\gamma), \lambda_{A-1}(\gamma), ..., \lambda_{A-k+1}(\gamma) \}$ on the main diagonal and on the upper diagonal, contiguous to the main diagonal, the positive interaction rates in $\boldsymbol{\lambda}$ except the last one (all rows add up to zero except the last row). 

Equation~\ref{eq:isec_density} can be written as a combination of the base exponential distributions, as in Eq.~\ref{eq:isec_density_distinct}, which is particularly useful for comparisons to other cascades. Since the interaction matrix $\boldsymbol{\Lambda}$ is upper triangular, it is nonsingular (provided all diagonal rates are different) and diagonalizable. Its diagonalized form $\boldsymbol{D}_{\Lambda} = \boldsymbol{J}^{-1} \boldsymbol{\Lambda} \boldsymbol{J}$ has the same diagonal elements as $\boldsymbol{\Lambda}$ (where $\boldsymbol{J}$ is an invertible matrix). Thus, Eq.~\ref{eq:isec_density} can be written as
\begin{equation}
    f^{\mathrm{IS}}_{A \to A-k}(L)= -\boldsymbol{b} e^{\boldsymbol{D}_{\Lambda} L } \boldsymbol{D}_{\Lambda} \boldsymbol{d} .
    \label{eq:isec_density_compare}
\end{equation}

The starting vector $\boldsymbol{b} = \boldsymbol{\phi} \boldsymbol{J}$ and the ending vector $\boldsymbol{d}=\boldsymbol{J}^{-1} \boldsymbol{1}$ depend on the contents of $\boldsymbol{\Lambda}$ and the central term $e^{\boldsymbol{D}_{\Lambda} L } \boldsymbol{D}_{\Lambda}$ has a diagonal form and is common to all interaction matrices, $\boldsymbol{\Lambda}$, having the same diagonal elements. Thus, it is useful for comparing cascades with the same total interaction rates but with a different number of channels. Equation\,\ref{eq:isec_density_compare} implies a linear combination of exponentials with rates from $\boldsymbol{\lambda}_{A \to A-k}$ and coefficients $c_k = -b_kd_k$ given by the elements of the starting and ending vectors. In this form, the physical meaning of the starting and ending vectors is lost and the coefficients $c_k$ may take complex values.

The expression for the distribution function of ISeCs is 
\begin{equation}
    F^{\mathrm{IS}}_{A \to A-k}(L)= 1 - \boldsymbol{\phi} e^{\boldsymbol{\Lambda} L} \boldsymbol{1} .
    \label{eq:isec_distribution}
\end{equation}
Some moments of interest are listed in Table~\ref{tab:moments}. In the cases where analytic expressions for the moments and variance are not available, some bounds can be established~\citep{He2019,He2021}. The distribution functions for an arbitrary mixture can be computed as in Eq.~\ref{eq:canonical_mixture}, where the distribution for each individual cascade has the form of Eq.~\ref{eq:isec_density}. However, it is more convenient to build the starting vector with the initial fractions, $\boldsymbol{\phi}_{\mathrm{mix}} = (\eta_A, \eta_{A-1}, ...., \eta_{A-k})$, and substitute in Eq.~\ref{eq:isec_density},
\begin{equation}
    f^{\mathrm{IS, mix}}_{A \to A-k}(L)= \boldsymbol{\phi}_{\mathrm{mix}} e^{\boldsymbol{\Lambda} L} \boldsymbol{\Lambda} \boldsymbol{1} .
    \label{eq:isec_mixture_density}
\end{equation}

\subsection{Concurrent cascades}
\label{subsec:concurrent_cascade}

The general cascade requires the inclusion of multiple channels at each step, producing a network of states. Unlike serial cascades where the path between any pair of states is unique, in the general case, each node may branch into multiple options forming a network of intersecting ISeCs that develop concurrently. These cascade types are referred to as concurrent cascades (CoCs) herein. One of the simplest examples in the literature is the disintegration scheme proposed by \cite{Puget1976}, the PSB model. In the PSB model, while there is only one species for each mass, each nucleus can jump to multiple other nuclei due to the additional disintegration channels, such as one- and two-nucleon emission in the GDR region and 6-15 nucleon emission in the quasi-deuteron region.
The density function for the distance until absorption is the same as in Eq.~\ref{eq:isec_density}, but the matrix $\boldsymbol{\Lambda}$ has additional terms in each row representing jumps to other nuclei in the chain. This is unlike the matrix for ISeCs, which contains only jumps to the immediate species with lower mass. An expression in the form of Eq.~\ref{eq:isec_density_compare} may not exist in general for CoCs as no set of coefficients $c_k$ can produce the equivalent function (see Appendix~\ref{sec:expected_distance}).

In their most general form, CoCs should include all known nuclear species, having multiple nuclei with the same mass number. The density is described as in Eq.~\ref{eq:isec_density} although the interaction matrix $\boldsymbol{\Lambda}$ and the starting vector $\boldsymbol{\phi}$ contain a larger number of rows, matching the increased number of species. The nondiagonal elements of the matrix $\boldsymbol{\Lambda}$ in this case are expressed as 
\begin{equation}
    \lambda_{S_i \to S_j} = \lambda^{\mathrm{tot}}_{S_i \to S_j}(\gamma) = \sum_k \lambda^k_{S_i \to S_j} (\gamma) + \gamma /c \sum_m \tau_m
    \label{eq:inclusive_rates}
\end{equation}
which denotes all $k$ photonuclear channels where species $S_i$ produces species $S_j$, and all $m$ decays having a decay time $\tau_m$, where $S_i$ decays into $S_j$. The sequence of indices $i, j$ in $\boldsymbol{\phi}$ and $\boldsymbol{\Lambda}$ is chosen in order of descending mass and charge numbers, as for RSeCs and ISeCs. This ensures that the matrix $\boldsymbol{\Lambda}$ is upper triangular, since disintegrations can only produce species with lower masses\footnote{The computation of the moments is more efficient when the matrix $\boldsymbol{\Lambda}$ is upper triangular, since fast computation methods are available (e.g. the Matrioshka matrices~\citep{Daw2019} apply).}. 
However, the lower triangular section of the matrix $\boldsymbol{\Lambda}$ may contain nonzero elements if there are nuclear decays that preserve the mass number, while increasing the charge number (e.g., $\beta^-$ decays). The main diagonal of the interaction matrix contains the total interaction rate for each species, $S_i$, which is the sum of all processes that lead to any other species, $S_j$, in the disintegration cascade, 
\begin{equation}
  \lambda_{S_i} = \lambda^{\mathrm{tot}}_{S_i}(\gamma) = \sum_{S_j} \lambda^{\mathrm{tot}}_{S_i \to S_j} (\gamma) .
  \label{eq:total_rates}
\end{equation}
With these elements, the resulting interaction matrix takes the form
\begin{equation}
    \boldsymbol{\Lambda}(\gamma) = 
    \begin{pmatrix}
        -\lambda_{S_1} & \lambda_{S_1 \to S_2} & \lambda_{S_1 \to S_3} & ... & \lambda_{S_1 \to S_N}\\
        0 & -\lambda_{S_2} & \lambda_{S_2 \to S_3} & ... & \lambda_{S_2 \to S_N}\\
        0 & 0 & -\lambda_{S_3} & ... & \lambda_{S_3 \to S_N}\\
        ... & ... & ... & ... &... &\\
        0& 0& 0& ... &-\lambda_{S_N}\\
    \end{pmatrix}
\end{equation}
where $N$ is the total number of species included, and the probability density and distribution functions for the distance until absorption are
\begin{equation}
    f^{\mathrm{CC}}(L)= -\boldsymbol{\phi} \exp \left( \boldsymbol{\Lambda} L\right) \boldsymbol{\Lambda} \boldsymbol{1}
    \label{eq:ph_density}
\end{equation}
\begin{equation}
    F^{\mathrm{CC}}(L)= 1 - \boldsymbol{\phi} \exp \left( \boldsymbol{\Lambda} L\right) \boldsymbol{1} .
    \label{eq:ph_distribution}
\end{equation}

\begin{figure*}[ht!]
\includegraphics[width=.442\textwidth]{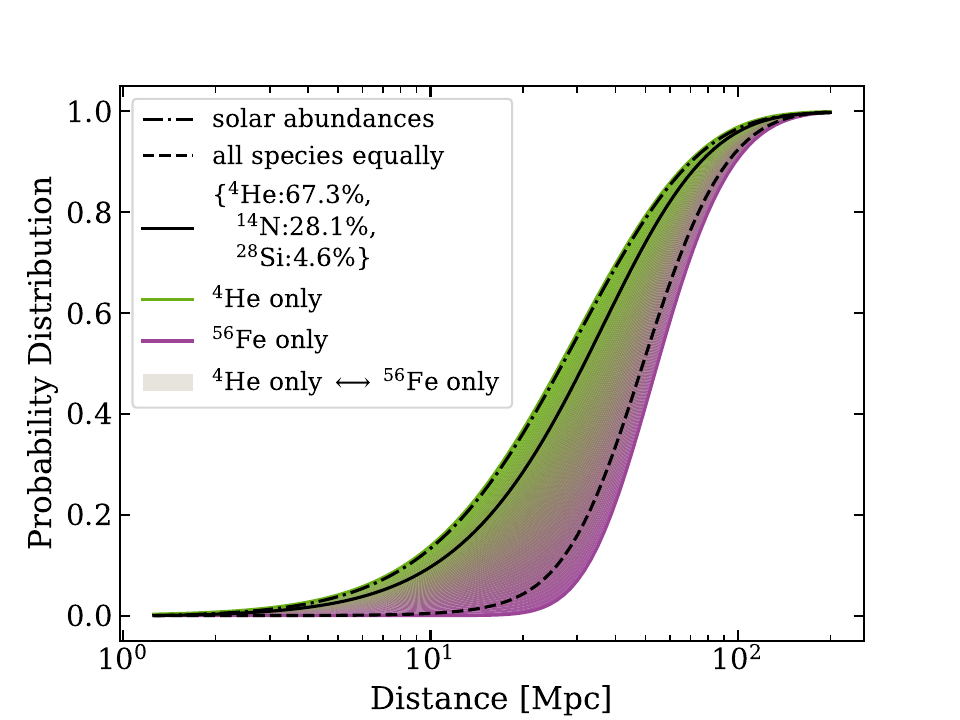}
\includegraphics[width=.558\textwidth, trim={.3cm 4.3cm 13.2cm 5.63cm}, clip]{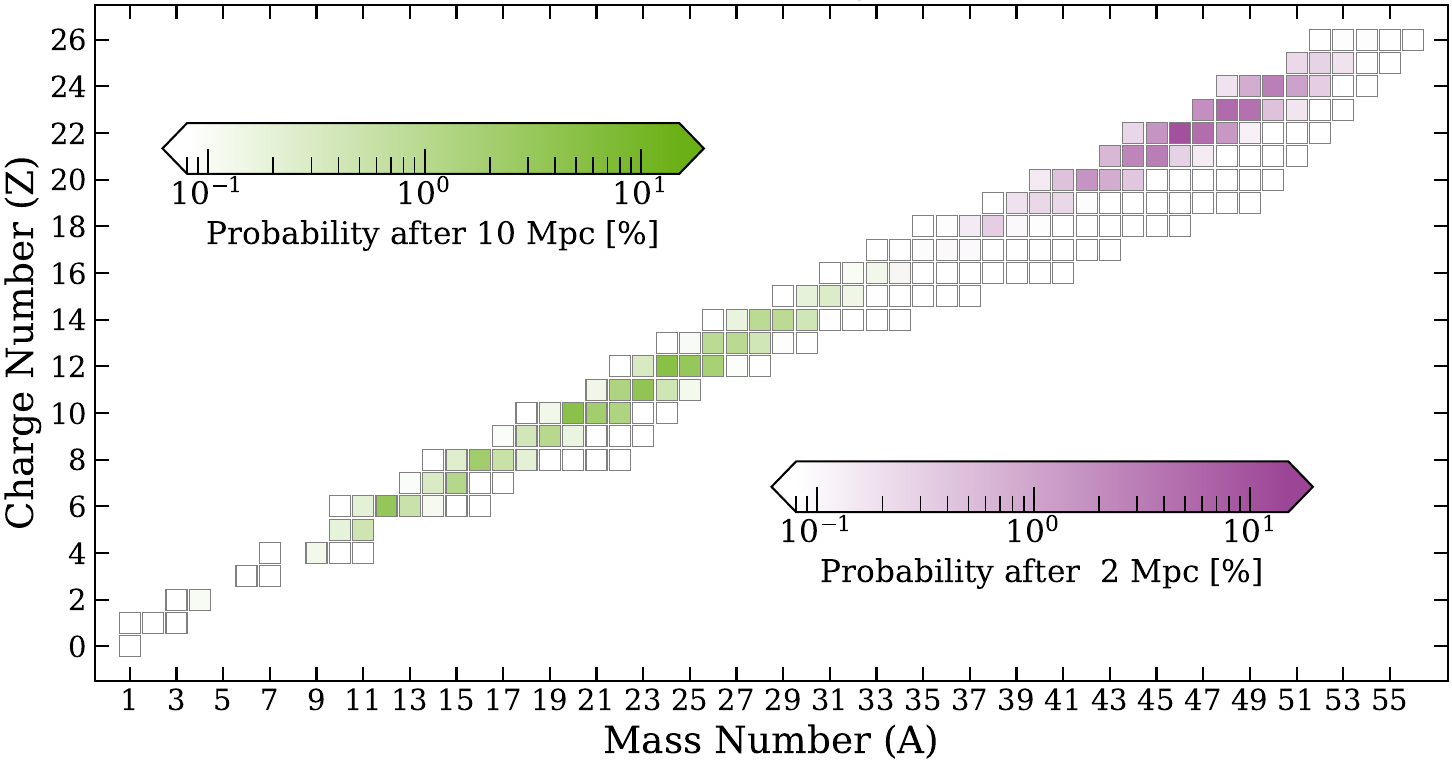}
\caption{Left: Probability of finding an injected nucleus or composition of nuclei fully disintegrated after propagating a specified distance ($\gamma=7\cdot10^9$). The green and purple solid lines in the extremes correspond to $^4$He and $^{56}$Fe injection, respectively, representing the lightest and heaviest initial compositions. 
The black solid line uses a similar composition as obtained in fits of the UHECR spectrum \citep{Abdul_Halim_2024}. The specific nuclei and their approximate fractions are given in the legend. The dashed black line represents the case where all species share the same fraction and the black dot-dashed black line a composition reflecting solar abundances. Right: Occupation probabilities for species in the nuclear cascade for $^{56}$Fe injection ($\gamma=7\cdot10^9$). The values are given for two propagation distances in the range where the full disintegration probability is negligible: 2\,Mpc (purple) and 10\,Mpc (green).}
\label{fig:mass_groups_distributions}
\end{figure*}

In CoCs, the "absorption state" may be a group of states (and not always a unique species) and it is represented by the absorption vector $\boldsymbol{\omega} = -\boldsymbol{\Lambda} \boldsymbol{1}$ whose components are the rates of transitioning to absorption from each of the species. For instance, when computing transitions between mass groups, $\boldsymbol{\phi}$ would contain nonzero values for nuclei with a mass equal to the injection mass number and the absorption vector $\boldsymbol{\omega}$ would be nonzero for nuclei with a mass equal to the final mass. Thus, this formulation allows us to study any possible type of cascade, and the construction of the matrix $\boldsymbol{\Lambda}$ encodes also the absorption state.

Figure~\ref{fig:mass_groups_distributions} (left) exemplifies these cascades by showing the probability that specific primary nuclei or a composition of primary nuclei with a Lorenz factor of $\gamma=7\cdot10^9$ have fully disintegrated after propagating a certain distance. The green and purple outer lines represent primary $^4$He and $^{56}$Fe, respectively.

Additionally, three examples of mixed composition are shown: solar abundance (black dot-dashed line), which is very light but contains a nonzero fraction of species heavier than helium; a UHECR-like composition (black solid line), which has elemental fractions similar to those obtained by fitting the UHECR spectrum and composition; and an equal fraction for all species (dashed black line) which places a larger fraction on heavier species because they are more numerous.

The figure illustrates the marked differences in composition evolution over typical UHECR propagation lengths ranging from 1 to 100\,Mpc. After a propagation distance of 100\,Mpc, all nuclei are fully disintegrated in the three cases, but the lighter mixtures reach 50\% probability of complete disintegration within 20-30 Mpc, while the heavier mixtures require 2-3 times larger distances. 

It should be noted that these distributions do not describe the occupation probabilities of species in the nuclear cascade, but only the probability of having the initial species fully disintegrated. Figure~\ref{fig:mass_groups_distributions} (right) represents such occupation probabilities for an initial $^{56}$Fe ($\gamma=7\cdot10^9$) after two different propagation distances for which the probability of full disintegration is almost null. After 2\,Mpc the average mass is reduced by 8-10 nucleons, given the short interaction lengths for iron and nuclei of similar mass. It might seem counterintuitive that after five times larger distance (10\,Mpc) the average mass is still $\sim20$ instead of five times the loss for 2\,Mpc, but this is the expected result given the reduction of the interaction lengths as the cascade moves to lower masses. The details of the cascade evolution demonstrate that injecting certain species as surrogates for mass groups is not a valid simplification for probability distributions.
\begin{figure}[t]
\includegraphics[width=.5\textwidth]{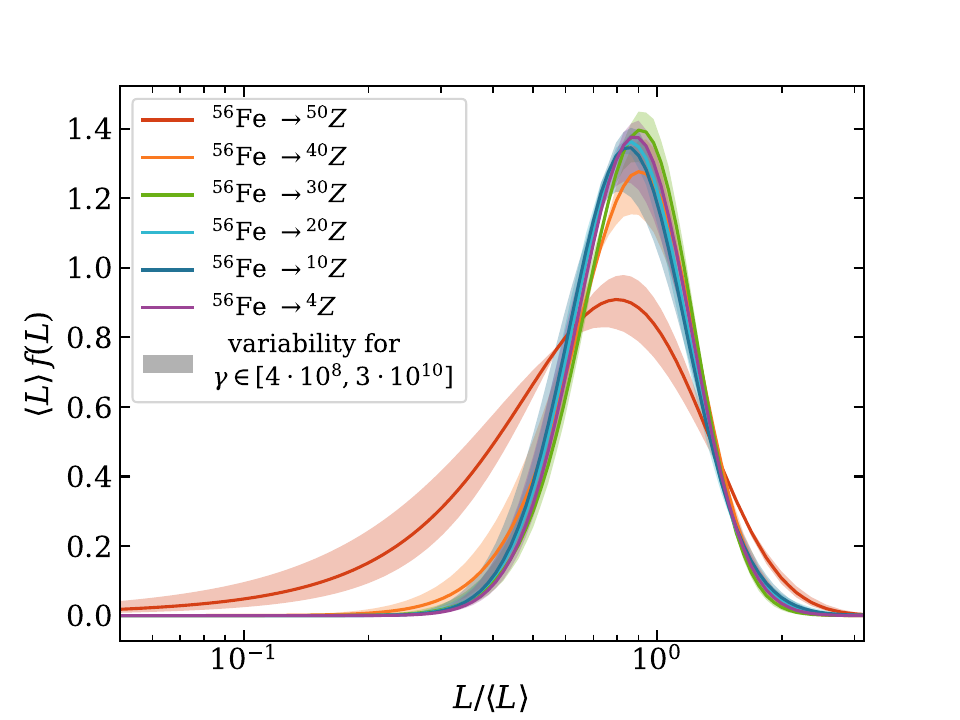}
\caption{Density functions of distance until reaching different values of nuclear mass, the variation for the boost $\gamma \in [4\cdot10^8, 3\cdot10^{10}]$ is represented by the shaded bands. The distributions are standardized and centered at the expected value, as they span different scales at different boosts.}
\label{fig:Comparing_distrbutions_by_abs_mass}
\end{figure}

Nevertheless, the correlation between a certain mass loss and the corresponding distance needed is remarkably stable. Figure~\ref{fig:Comparing_distrbutions_by_abs_mass} illustrates this relation over a broad boost range. Since the distributions span from a few to thousands of megaparsecs depending on the boost, the mean of the distribution was used to regularize the distance scale. The density distributions for the distance until the initial state $^{56}$Fe is absorbed into a certain mass group (indicated with different colors) are shown with solid lines representing the mean, and shaded bands bracketing the extreme values at each distance point, as the boost moves in the range $4 \cdot 10^8$ to $3 \cdot 10^{10}$. The remarkable regularity of the distributions is evident from the negligible variation, especially considering the broad range of length scales and the differences in the target photon fields. Indeed, $\langle L \rangle$ is in the sub- to few megaparsec scales for $\gamma \gtrsim 3 \cdot 10^9$ (predominantly CMB interactions) and in the tens to gigaparsec scales for $\gamma \lesssim 3 \cdot 10^9$ (predominantly IRB interactions). Distributions involving only a few species have a broader relative width, as seen for absorption at mass 50, becoming narrower with the increase of intermediate species, and showing little change for absorption masses below 40. A discussion of the implications of this regularity in extragalactic propagation is included in Sect.~\ref{sec:atrophysical_impact}.

\subsection{Light secondary products}
\label{subsec:secondaries}

In addition to the leading mass, nuclear cascades produce multiple light nuclei, such as deuterium and $\alpha$-particles, which can be considered boost-preserving products. They also produce light secondaries, such as pions and single nucleons, which are produced with a broad spectrum of energies. Larger nuclear fragments may also be present. For example, photo-fission yields at least two fragments of similar mass. In this stochastic description, all of these products are treated as additional particles in each stochastic jump. In the discussions above, the largest mass nucleus has been used as the nominal species, denoting the current state of the cascade. Here, we describe the treatment of the aforementioned secondary products, which are produced during state jumps in the cascade. 

Clearly, the production of light secondaries is also stochastic, as it relates to transitions in the developing cascade. A simple approach to compute the production of the $k$-th secondary, $\frac{d}{dL}Q_k(\gamma, L)$, as a function of the path length $L$ and the Lorentz boost $\gamma$
\begin{equation}
    \frac{dQ_k}{dL}(\gamma) = \phi \frac{d}{d L} \boldsymbol{P}^L \boldsymbol{Y}_k \boldsymbol{1} = \phi \boldsymbol{P}^L(L, \gamma) \boldsymbol{\Lambda}(\gamma) \boldsymbol{Y}_k(\gamma) \boldsymbol{1} ,
    \label{eq:lights_production}
\end{equation}
where the yield matrix $\boldsymbol{Y}_k(\gamma)=\{y^k_{ij}(\gamma)\}$ contains the number of light secondaries of species, $k$, produced in jumps from any species $j$ to $i$. The $\boldsymbol{Y}_k$ matrix is strictly lower triangular, although some of the upper triangular elements could be nonzero, as discussed for the lower triangular part of the matrix $\boldsymbol{\Lambda}$. 

Boost-preserving products are injected into the same boost. For products with a broad spectrum, the boost distribution is described by a function $dn^k_{i \to j}/dx$, where $x$ is the fraction to the primary energy, $x=E_k/E_i \approx A_k/A_i \gamma_k/\gamma_i$ (generally independent of the boost). The norm is equal to the yield $y^k_{ij} = \int dn^k_{i \to j}/dx $. The treatment of the production of these light particles is well understood~\citep{2010ApJ...721..630H,Morejon2019} and the evolution of their spectrum over propagation can be computed analytically~\citep{Berezinsky:1990qxi}.

\section{Continuous energy losses}
\label{sec:continuous_losses}

The stochastic processes discussed so far do not account for the effect of CELs, which are deterministic (nonstochastic) interactions that cause energy losses without altering the nuclear species. This degradation in energy affects the Markov property of the cascade because the rates are no longer constant, but evolve as the Lorentz boost changes. Thus, CELs correspond to inhomogeneous continuous-time Markov chains, which violate the time homogeneity (the temporal independence of the rates of jumps between states). This means that the current state of the cascade depends on the entire past history rather than just the previous state.

In our context, it is useful to distinguish between two types of inhomogeneities caused by CELs: coherent inhomogeneities (CI), in which the present state depends on the total time (distance) elapsed, but not on the specific history of the process (i.e., the sequence of species), and dispersive inhomogeneities (DI), where the probability of the present state depends on the specific sequence of species in the past history. The latter type (DI) leads to differences in the boost evolution of the underlying concurrent cascades (dispersion), whereas in the former (CI) all concurrent cascades experience the same boost evolution (coherence). The effects of CI can be accommodated analytically through variable transformations if the time-dependence of the CI is known. DI effects are generally not analytically computable, but approximations and numerical methods are available for such cases \cite[e.g.,][]{Arns2010}. The relevant scenarios are discussed below for both the propagation of UHECRs and in-source interactions.

\subsection{Coherent inhomogeneities}
\label{subsec:coherent_inhomogeneities}

Cases of coherent inhomogeneities involve target photon fields that vary over time, since all rates are  affected by the time-dependence of the field regardless of the state of the cascade. For sources, a fireball scenario fits this description, given the adiabatic cooling of the interaction volume as it expands. In the case of propagation, the cosmological evolution of the target photon backgrounds and the adiabatic losses experienced by UHECRs can produce such inhomogeneities.

In general, when the interaction rates can be expressed as the product of a scaling function $\mu(L)$ dependent on distance (or redshift, time, etc.) and a rate dependent on the boost, the distribution and the density functions are analogous to the homogeneous ones \citep{Albrecher2019, Zhang2020}:
\begin{equation}
    f(L)= -\mu(L) \boldsymbol{\phi} \exp \left( \int_0^L \mu(s) ds \boldsymbol{\Lambda}\right) \boldsymbol{\Lambda}\boldsymbol{1}
    \label{eq:iph_density}
\end{equation}
\begin{equation}
    F(L)= 1 - \boldsymbol{\phi} \exp \left( \int_0^L \mu(s) ds \boldsymbol{\Lambda}\right) \boldsymbol{1} .
    \label{eq:iph_distribution}
\end{equation}
For example, suppose the target photon density is a function of time of the form $n(\epsilon, t) = m(t)n_0(\epsilon)$. The corresponding rates after integrating Eq.~\ref{eq:interaction_rate} take the form $\lambda(\gamma, t) = m(t)\lambda_0(\gamma)$, the interaction matrix constructed using the rates $\lambda(\gamma, t)$ results in a product of a time dependent scalar and a time independent matrix $\Lambda(\gamma, t) = m(t)\Lambda_0(\gamma)$, and Eqs.~\ref{eq:iph_density}-\ref{eq:iph_distribution} apply with $\mu(s) \equiv m(s/c)$. Comparing Eqs.~\ref{eq:iph_density}-\ref{eq:iph_distribution} to Eqs.~\ref{eq:ph_density}-\ref{eq:ph_distribution} makes it clear that they are equivalent if the propagated length in Eqs.~\ref{eq:iph_density}-\ref{eq:iph_distribution} is understood as the target thickness traversed $\delta =  \int_0^L \mu(s) ds$, and the expressions become identical when no scaling occurs, as $\mu(s)=1$ produces $\delta \equiv L$. 
\begin{figure*}[t]
\includegraphics[width=\textwidth]{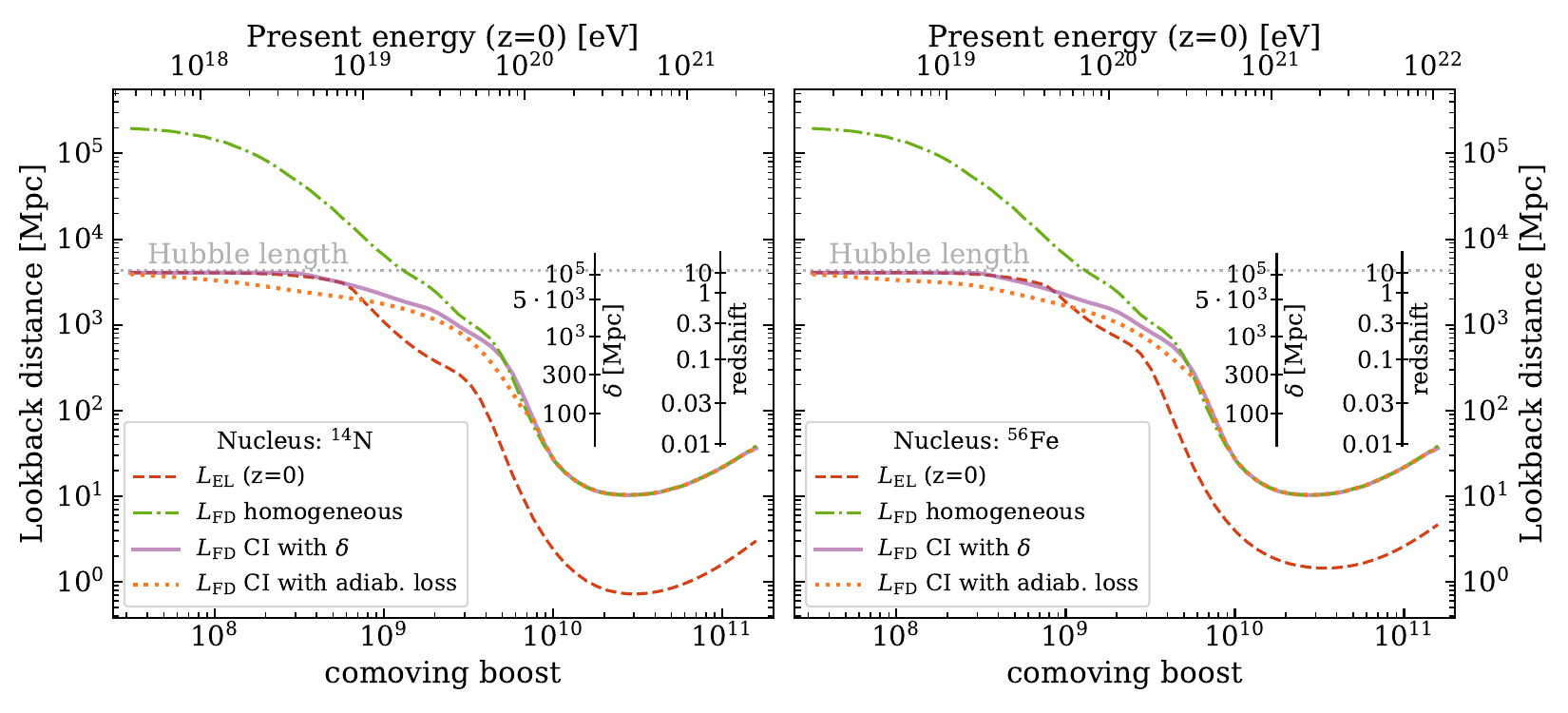}
\caption{Cosmic ray horizons of $^{14}$N (left) and $^{56}$Fe (right) in the background photon fields. The widely used energy loss length (dashed red) overestimates the effect of interactions. The $L_{\rm FD}$ horizons correspond to the distance where the full disintegration probability is 99~\%: in dot-dashed green the values for the homogeneous case, in solid purple the values assuming coherent inhomogeneities, and in dotted orange the values obtained numerically treating the redshift dependence of the rates and adiabatic losses.}
\label{fig:interaction_legths}
\end{figure*}
The application to source scenarios is evident when the interaction region expands adiabatically. In these cases, the geometry of the volume informs the functional dependence of $m(t)$, which governs the scaling of the target photon field. Similarly, for plasmoids moving along jets the scaling of the external photon fields could result in a change of only the norm \citep[e.g.,][]{Hoerbe2020}, in which case the temporal evolution would determine the form of $m(t)$. Examples of time dependence in a GRB are shown in Sect.~\ref{subsec:sources}
.

In the case of extragalactic propagation, the redshift scaling of the photon densities for the CMB and IRB leads to the convenient form for the interaction rates,
\begin{equation}
\lambda(\gamma, z) = a(z)(1+z)^3 \cdot \lambda((1+z)\gamma, z=0) ,
\label{eq:rates_on_z_g} 
\end{equation}
using the scaling prescription of \cite{Kampert:2012fi}, where $a(z)$ reflects the ratio of the photon number density at redshift $z$ and to the present one ($a(z) \equiv 1$ for the CMB). The redshift-dependent argument $(1+z)\gamma$ of the rates in Eq.~\ref{eq:rates_on_z_g} produces an additional boost drift in the rates so they are not truly separable into a redshift-dependent and boost-dependent components. However, the volume compression factor plays a larger role and the argument $(1+z)\gamma$ can be considered constant for sufficiently small propagation lengths\footnote{For example, for a cosmic ray starting 300~Mpc from us ($z\approx 0.073$) its Lorentz boost changes by only 7.3~\% as it propagates to Earth, while the initial interaction rate is $(1+z)^3\approx1.24$ is 24~\% greater than at present and $\delta\approx334$~Mpc, about 11~\% greater than the propagation distance.}. With this assumption, the interaction matrix can be written as a constant matrix multiplied by $a(z)(1+z)^3$, yielding a thickness of
\begin{equation}
    \delta = \int_0^L a(z)(1+z)^3 ds = \int_{z(0)}^{z(L)} a(z) \frac{c(1+z)^2}{H(z)} dz    .\label{eq:cosmological_thickness}
\end{equation}
This thickness has units of distance and can be interpreted as the equivalent propagation distance that a cosmic ray would need to cross to experience the same number of interactions as produced by the increased photon density. The distributions for extragalactic propagation assuming only the thickness is affected by the redshift are analogous to the homogeneous ones
\begin{equation}
    f^{\rm CI}(\delta)= -\boldsymbol{\phi} \exp \left( \boldsymbol{\Lambda}(\gamma)\delta \right) \boldsymbol{\Lambda}(\gamma)\boldsymbol{1} ,
    \label{eq:iph_eg_density}
\end{equation}
\begin{equation}
    F^{\rm CI}(\delta) = 1 - \boldsymbol{\phi} \exp \left( \boldsymbol{\Lambda}(\gamma)\delta \right) \boldsymbol{1} ,
    \label{eq:iph_eg_distribution}
\end{equation}
with the cosmological thickness replacing the propagation distance and noting that
\begin{equation}
        f^{\rm CI}(\delta) = \frac{d}{d\delta}F^{\rm CI}(\delta) = \frac{1}{a(z)(1+z)^3}\frac{d}{dL}F^{\rm CI}(\delta) .
\end{equation}

Figure~\ref{fig:interaction_legths} compares propagation horizons for $^{14}$N (left) and $^{56}$Fe (right) on the basis of the disintegration. The abscissa uses the comoving boost $\gamma_c=\gamma / (1+z)$, because it does not change during propagation under CI. For reference to previous works, the energy loss length, 
\begin{equation}
        L_{\rm EL}=\sum_j \frac{E_i}{(E_i - E_j) \lambda_{Ai \to A_j}(\gamma_c)} \approx \sum_j \frac{A_i}{(A_i - A_j) \lambda_{Ai \to A_j}(\gamma_c)}
\end{equation}
has been included (with present interaction rates, $z=0$), as it is often used to estimate the propagation horizons. The continuous limit is implicitly assumed in $L_{\rm EL}$ as it estimates the distance required to dissipate the initial energy, $E_i$, at an average energy loss rate computed from the interaction channels of the initial species. The horizons derived here take into account the stochastic nature of the process by including the probability that the initial species has fully disintegrated. The distance  $L^{\ell}_{\rm FD}$ at which the full disintegration distribution $F_{\rm FD}( L^{\ell}_{\rm FD}) =\ell$, reaches a desired limit $\ell$, constrains the probability of not fully disintegrating to $1-\ell$. The nominal values employed here take $\ell=99\%$ to define $L_{\rm FD} = L^{99\%}_{\rm FD}$ as the full disintegration horizon. The various lines for $L_{\rm FD}$ correspond to the different approaches employed. The dot-dashed-green line employed the distribution of the homogeneous case, where cosmological effects are ignored; hence, the values of lookback distance larger than the Hubble length. The solid-purple line employed Eq.~\ref{eq:iph_eg_distribution}, where the coherent inhomogeneities are taken into consideration only in the thickness, neglecting the redshift dependence of the second term in Eq.~\ref{eq:rates_on_z_g}. The dotted-orange values were computed by numerically integrating Kolmogorov's differential equation and updating the boost at each step to reflect the redshift dependence of the second term in Eq.~\ref{eq:rates_on_z_g} and also adiabatic losses. Additional grids denote the initial energy of the cosmic ray (top $x$-axis), the redshift corresponding to the lookback distance (rightmost inset scale, using a flat $\Lambda$-CDM cosmology with values fitted to the WMAP data, \cite{Bennett_2013}) and the thickness corresponding to the lookback distance (left of the redshift scale, assuming $a(z) = 1 $). 

The cosmological effects are negligible for distributions spanning a few hundred  megaparsecs and for Lorentz boosts where the CMB is the predominant photon target ($\gamma_c \gtrsim 5 \cdot 10^9$) as reflected in the identical values of $L_{\rm FD}$ for all approaches with and without cosmological effects. In this range of boosts, $L_{\rm EL}$ yields much shorter horizons than $L_{\rm FD}$, even considering the widths of the probability distributions. This is not surprising as $L_{\rm EL}$ assumes that the average energy loss rate is comparable to the initial values, neglecting the rate decrease as the mass of the cascade decreases and the stochastic fluctuations of the interaction lengths. For lower boosts ($\gamma \lesssim 5 \cdot 10^9$) interactions with IRB photons become dominant, and cosmological effects become appreciable in the separation of the homogeneous horizons from the CI horizons. The solid-purple line includes the cosmological effects only on the thickness, so it illustrates the difference between $\delta$ and the lookback distance as it comes from evaluating the homogeneous distribution on $\delta$ instead of the lookback distance as in the dot-dashed-green line. The differences between these two curves quantify the error of employing the lookback distance instead of the thickness, and the limitations of the homogeneous approach. The comparison of the thickness and the lookback distance scales is sufficient to decide which approach is needed. 

The dotted-orange line includes, besides the thickness increase present in the purple-solid line, the boost shifts produced by adiabatic losses and the cosmological energy increase of the photon backgrounds. We note that while the physical boost changes, the comoving boost, $\gamma_c$, does not change in the case of CI (see Sect.~\ref{subsec:dispersive_inhomogeneities}). The shorter horizons below $\gamma \lesssim 5 \cdot 10^9$ are a consequence of the boost increase with redshift, which probes larger values of the interaction rates as they are monotonically increasing with the boost. Thus, the inclusion of $\delta$ alone is not sufficient to describe the cosmological effects, but it is a reasonable upper limit with an error appreciable here by comparing the solid-purple and the dotted-orange curves.
The present description confirms the so called ``explosive regime'' in the mass evolution observed by \cite{Aloisio2013AnalyticRadiation} using a continuous approach and the one-nucleon loss approximation. We demonstrate this is also a feature of the stochastic cascades produced by the increase of the thickness with lookback distance, which becomes very large as redshifts above 0.1. Naturally,  this is a general property of UHECR interaction cascades during propagation, irrespective of the nuclear interaction model and the target photon field (see Appendix~\ref{sec:di_decoherence_lengths}).

\subsection{Dispersive inhomogeneities}
\label{subsec:dispersive_inhomogeneities}

Energy losses that depend on the nuclear species affect the cascade development in variable degrees depending on the specific sequence of states, thus the total energy loss after multiple disintegrations can vary significantly among the concurrent disintegration chains. This implies that different sequences within CoCs could produce diverging evolutions of the Lorentz boost , thus gradually rendering the cascade incoherent. 

Examples of CELs that cause DIs include: synchrotron losses, which are relevant within sources with strong magnetic fields; and pair production losses, which are relevant for extragalactic propagation. The rate at which the boost changes (equivalent to the energy loss rate) for synchrotron losses is
\begin{equation}
    -\frac{1}{\gamma}\frac{d\gamma}{dL} = \frac{\sigma_T m_e^2}{6\pi m_p^4} \gamma B^2 \left(\frac{Z}{A}\right)^4
\end{equation}
where $\sigma_T$ is the Thomson cross-section, $m_e$ and $m_p$ are the masses of electrons and protons, respectively, and $B$ is the magnetic field intensity in the source. The relation for the boost change in this expression depends on the nuclear species, given the factor $\left(\frac{Z}{A}\right)^4$. Thus, the losses are affected by the specific sequence of nuclei and the distances traveled by each nucleus.

The rate of boost change for pair production losses \citep{Blumenthal1970} is
\begin{equation}
    -\frac{1}{\gamma}\frac{d\gamma}{dL} = \alpha r_0 m_e^2c^4 \frac{Z^2}{\gamma A} \int_2^\infty d\xi \, n\left(\frac{\xi m_ec^2}{2\gamma}\right) \frac{\phi(\xi)}{\xi^2} .
\end{equation}
This expression is also dependent on the nuclear charge and mass numbers. Following the notation $\beta_0 / c = \frac{1}{\gamma^p}\frac{d\gamma^p}{dL}$ for the loss length of protons \citep{ALOISIO201373}, the losses of nuclei in general can be written as
\begin{equation}
    \frac{1}{\gamma}\frac{d\gamma}{dL} = \frac{Z^2}{A} \frac{\beta_0(\gamma)}{c} .
    \label{eq:boost_evol_present}
\end{equation}
In the previous section, we discuss how adiabatic losses can be treated as CI, as well as the redshift scaling of the rates produced by the corresponding scaling of photon backgrounds . The complete form of the boost evolution with the redshift dependencies and  adiabatic losses has been obtained for the kinetic equation formalism~\citep{ALOISIO201373} via
\begin{equation}
    \frac{1}{\gamma}\frac{d\gamma}{dL} -\frac{1}{1+z} \frac{dz}{dL}= a(z)(1+z)^3 \frac{Z^2}{A} \frac{\beta_0((1+z)\gamma)}{c} ,
    \label{eq:eprod_cel_cosmo}
\end{equation}
where we can consider $a(z)=1$ since pair production losses are dominated by the CMB. Rewriting Eq~\ref{eq:eprod_cel_cosmo} in terms of the comoving Lorentz boost and the thickness, we obtain the principal relation that governs boost changes for cosmological propagation,
\begin{equation}
    \frac{1}{\gamma_c}\frac{d\gamma_c}{d\delta} = \frac{Z^2}{A} \frac{\beta_0 ((1+z)^2\gamma_c)}{c} .
    \label{eq:comboost_evol}
\end{equation}
The CI approach corresponds to the homogeneous form of this differential equation, since in the CI approach $\gamma_c$ is constant, as obtained for the solution in the absence of CELs (right-hand side null). Thus the effect of pair production losses is a modification of the CI result caused by a drift in the comoving boost.

The evolution of the boost over cosmological thickness can be obtained by integrating numerically Eq.~\ref{eq:comboost_evol}, or via a variable separation for sufficiently short propagation distances, such that $z$ is almost constant,
\begin{equation}
    \int_{\gamma^1_c}^{\gamma^2_c} \frac{cd\gamma_c}{\gamma_c\,\beta_0 ((1+z)^2\gamma_c)} =  \Phi(\gamma_c^2)-\Phi(\gamma_c^1) = \frac{Z^2}{A} \delta ,
\end{equation}
where it is evident that the change in comoving boost from the initial $\gamma^1_c$ to a final $\gamma^2_c$ for any nuclear species is proportional to the thickness. For protons, this change is given by the function $\Phi(\gamma_c)$, which can be precomputed numerically for ranges of the redshift. For nuclei, the thickness required for a comparable change in the comoving boost is smaller by a factor of $Z^2/A$. 

The relevance of DI is limited to a small region of boost and redshift phase space relevant to the propagation of UHECRs (see Appendix~\ref{sec:di_decoherence_lengths}). Within this range, they can be accounted for using a quasi-homogeneous approach, in which the thickness is divided into segments small enough that DI are negligible, ensuring the CI description is sufficient. This is possible because, in any cascade, there is a dominant rate (typically for the species with the largest mass) and for sufficiently small values of $\delta$, the constancy of $\gamma_c$ can be assumed. The cascade can then be described by a set of CI descriptions, each applying within a segment in which the interaction matrix is evaluated at the constant boost of the segment. The boost values are updated after each segment by selecting the most likely value. The total boost change is also stochastic, but its distribution can be determined with a transformation via rewards~\citep{Bladt2017}.

\section{Astrophysical examples} 
\label{sec:atrophysical_impact}

\subsection{Distance horizons and mass evolution}
\label{subsec:mass_evolution}

The regularity of the mass evolution with distance, described in \cite{MorejonPhdThesis} as a disciplined disintegration (DD), was invoked to explain the gradual decrease of the average mass with propagation distance observed in CoCs computed with PriNCe~\citep{Heinze:2019jou}. The DD was derived assuming the regularity condition and only one-nucleon emission per interaction, which lead to the distance until a set nucleon-loss being proportional to the inverse of the mean interaction rate per nucleon and the logarithm of the ratio of initial and final masses \citep[see Eq. 6.11 in][]{MorejonPhdThesis}. This is consistent with the expectation value in Table~\ref{tab:moments} for RSeCs. The validity of DD in CoCs was not quantitatively verified in \cite{MorejonPhdThesis}, but rather inferred from a small number of simulations. We expand on this idea here by first considering ISeCs, which are serial but do not follow the regularity condition, and then considering the more general CoCs, which include multiple branching channels.

\begin{figure}[t]
    \centering
    \includegraphics[width=0.5\textwidth]{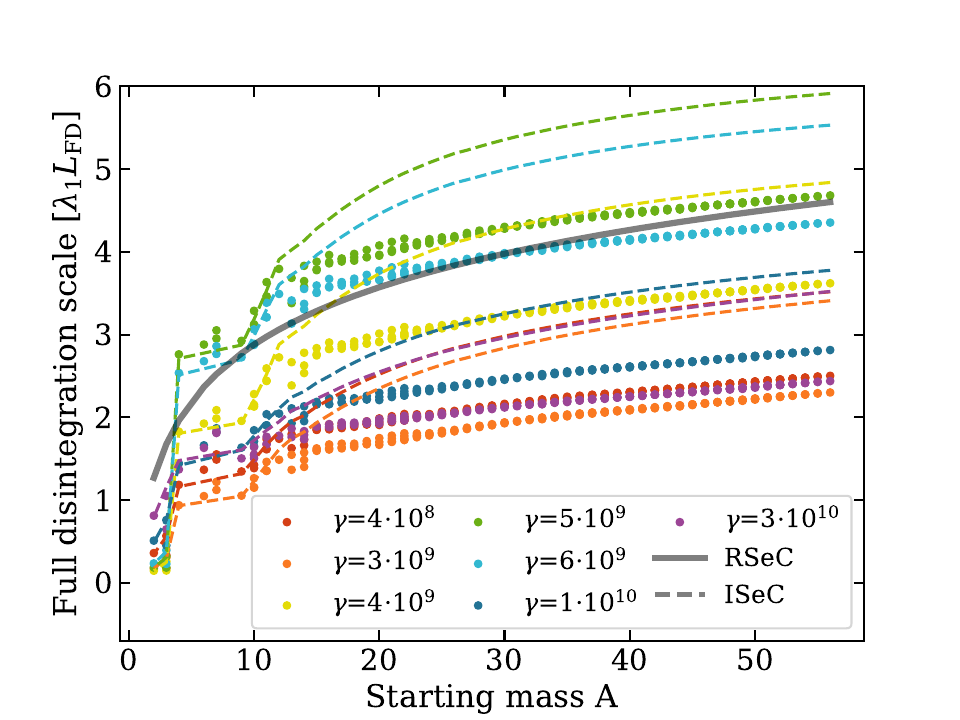}
    \caption{Expected distance until full disintegration in units of the inverse of the mean interaction rate per nucleon $1/\lambda_1$. The black line corresponds to the canonical cascade (RSeC), the dashed lines represent ISeCs, and the scattered points show values for CoCs, where multiple points appear for each mass corresponding the multiple isobars. The boost is indicated by the color, as listed in the legend.}
    \label{fig:fdd_vs_mass}
\end{figure}

Figure~\ref{fig:fdd_vs_mass} compares the relationship between the initial mass and the expectation of the full disintegration length ($L_{\rm FD}$, mean of the distribution) scaled by the mean interaction rate per nucleon $\lambda_1$. Different boost values contrast the changes in photodisintegrations as the rates transition from IRB-dominated to CMB-dominated with boost increase. The RSeC reference represents the relation $\lambda_1L_{\rm FD} = \ln A$ with a solid black line. The ISeCs, based on the total cross-sections in CRPropa~3.2~\citep{Kampert:2012fi,2022JCAP...09..035A}, are represented by dashed lines, with colors indicating the boost values listed in the legend. Since the ISeCs cover only one species per mass and only one-nucleon loss, the rates employed are an average over nuclei of the same mass, and all other channels present in the cross-section table are ignored. This ensures a consistent comparison with CoCs (dots), which are based on the same cross-section table but include all channels and multiple species per nuclear mass. 

\begin{figure*}[t]
    \centering
    \includegraphics[width=0.95\linewidth]{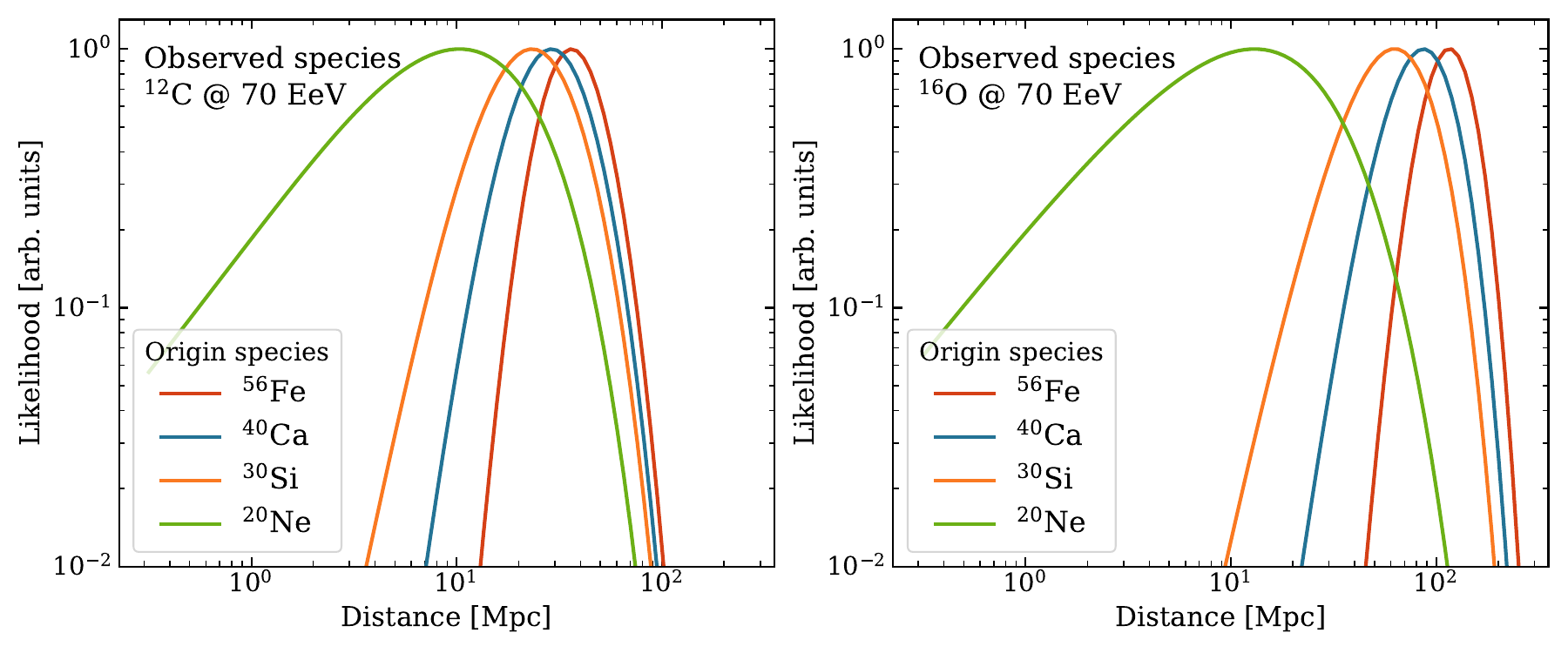}
    \caption{Likelihood of the distance of origin of an observed 70 EeV  $^{12}$C (left) and $^{16}$O nucleus assuming different initial nuclei. Even slight variations in the mass of the observed nuclei can lead to significant differences in their most likely distance of origin.}
    \label{fig:backpropagation}
\end{figure*}

Overall, ISeCs exhibit a similarly behavior to RSeCs, apart from a boost-dependent offset that can be attributed to the variance in the mean interaction rate per nucleon. The proportionality to $\ln A$ found in RSeCs is a consequence of the serial character that is also found in ISeCs. However, the irregularities in ISeC rates produce offsets in the mass dependence that can be as large as $3\left|1 - \frac{A_k\langle\lambda_1\rangle}{\lambda_{A_k}}\right|$ (c.f., Appendix~\ref{sec:expected_distance}), where $A_k$ is the mass of the species in the cascade for which the interaction rate, $\lambda_{A_k}$, deviates the most from the regular rate, $\langle\lambda_1\rangle A_k$. These offsets can also vary with starting mass, because additional species included in cascades of heavier masses can slightly contribute to the rate variability. However, the main impact comes from the dependence of the rates on the boost: at the lowest and highest boost values the offsets are comparable, and they increase at intermediate values. This progression is related to the onset of photodisintegrations with the CMB: in the boost region  $4\cdot 10^9-6\cdot 10^9$, the rates integrate the energy-weighted cross-section from $\varepsilon \lesssim 20$~MeV to $\varepsilon \lesssim 40$~MeV, where the variance among species is the largest (see Fig.~\ref{fig:rates_variations}, CRPropa). As the boost increases and the variance decreases, the offset values become comparable to those around $\gamma=4\cdot 10^8$, where interactions with the IRB dominate.  For masses lower than $A=12$, the trend is visibly disrupted, possibly due to limitations in cross-section data employed for these nuclei~\citep{Kampert:2012fi}.

The additional disintegration channels have a significant effect on the CoCs models, which include all possible nucleon losses in the cross-section table. Multiple dots in each mass correspond to the different isobars; however, their differences become negligible for $A\gtrsim 23$ as the number of concurrent cascades increases, thereby smoothing the isobar variance. CoCs exhibit a linear rather than logarithmic mass dependence, which is a clear sign that the multiple concurrent cascades enhance the efficiency of the disintegration, thereby shortening the length scales (see Appendix~\ref{sec:expected_distance}). Nevertheless, the proportionality of $L_{\rm FD}$ to the mass is why the DD effect holds in CoCs, as evidenced by PriNCe simulations~\citep{MorejonPhdThesis} at $\gamma=2\cdot 10^{10}$ for nuclei up to lead ($A=208$). However, the explanation proposed by \cite{MorejonPhdThesis} is incomplete and applies only to serial cascades; it fails to reproduce the linear behavior demonstrated here. 

The significant changes in length scales associated with boosts are a valuable feature that could be leveraged in future studies using the precise description proposed here and assuming the required accuracy in the cross-section data. Focusing on UHECRs in the boosts where CMB interactions begin to dominate, comparisons of events of adjacent boosts could allow probing different origins. Specifically, in the boost region $3\cdot 10^9-1\cdot 10^{10}$, the horizons end up shortened considerably (see Fig.~\ref{fig:interaction_legths}), while the full disintegration length scale can vary drastically between adjacent boosts. For example, comparing $3\cdot 10^9$ to $5\cdot 10^9$ ($\sim66$\% change) implies a reduction by more than half in $L_{\rm FD}$, for both light ($^{14}$N) and heavy nuclei ($^{56}$N). Additionally, dispersive inhomogeneities have a larger influence in this range (see Appendix~\ref{sec:di_decoherence_lengths}), which could enhance the differences between adjacent boosts. Extending the comparisons to slightly lower values, where IRB interactions still dominate, could allow testing the emitted spectrum in the paradigm of identical sources, as the expected changes in composition can now be computed with remarkable accuracy, including the stochastic effects and the probability distributions for individual events. In this paradigm, changes in composition for different energies would encode the relative contribution from different distances, since the observed composition can be efficiently computed with arbitrary precision. This allows us to employ the approach in minimization algorithms.

The verified DD effect implies that the cosmic ray horizon can be precisely defined as a quantity that naturally results from the photodisintegration cross-sections, the opacity of the target photon field, and the stochastic nature of cosmic ray propagation, rather than as an effective quantity dependent on source properties, such as the emission spectrum or the cosmic density. This quantity should be a function of the initial species, boost and redshift, such as the full disintegration horizon $L^{\ell}_{\rm FD}$ obtained from the distributions of distance until full disintegration, as discussed in Sect.~\ref{subsec:coherent_inhomogeneities}. Such a limit is meaningful even when considering magnetic deflections. The heaviest species in a composition has the largest horizon due to the DD effect, and their rigidity tends to be the largest. Indeed, $R = E/Z = \gamma/\kappa$ and the charge-to-mass ratio $\kappa = Z/A$  (typically within 0.3-0.6 for all nuclei and within 0.4-0.5 for stable nuclei) tends to be lower for heavier nuclei. Thus, the products of the heaviest nuclei emitted would propagate farther and experience the least magnetic deflections (see Sect.~\ref{subsec:magnetic_deflections}). The full disintegration limit constrains the propagation length, which is equivalent to the distance reached under ballistic propagation. However, under diffusive propagation, the distance reached by nuclei would be shorter because the lengths of diffusive paths tend to be larger than the radial distances reached. The effect of diffusive motion in sources is illustrated in Sect.~\ref{subsec:sources}. 

\begin{figure*}[t]
    \centering
    \includegraphics[width=0.49\textwidth]{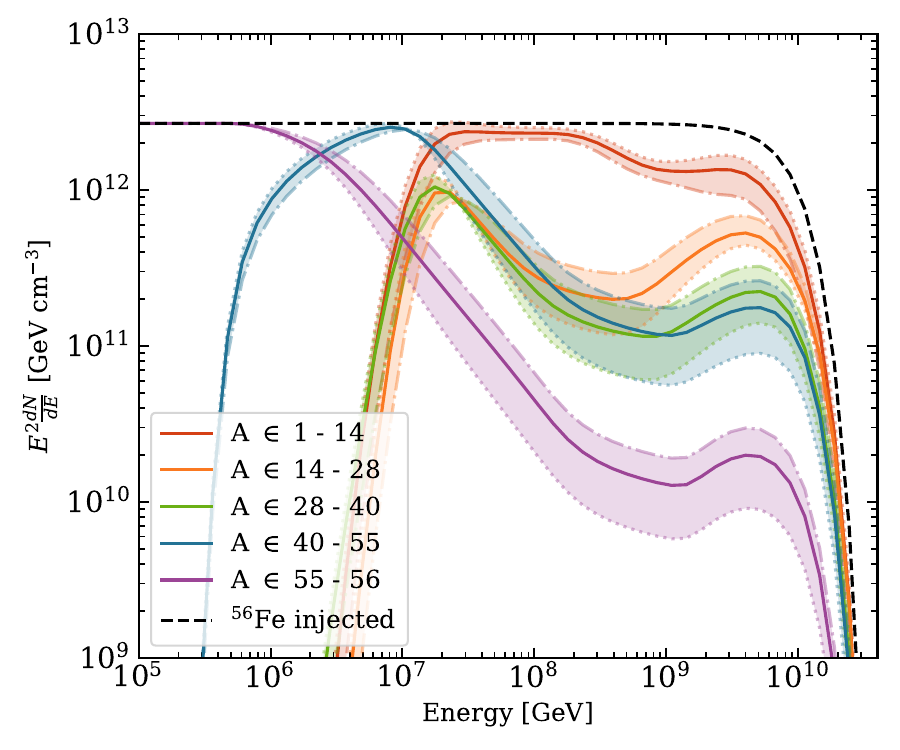}
    \hfill
    \includegraphics[width=0.49\textwidth]{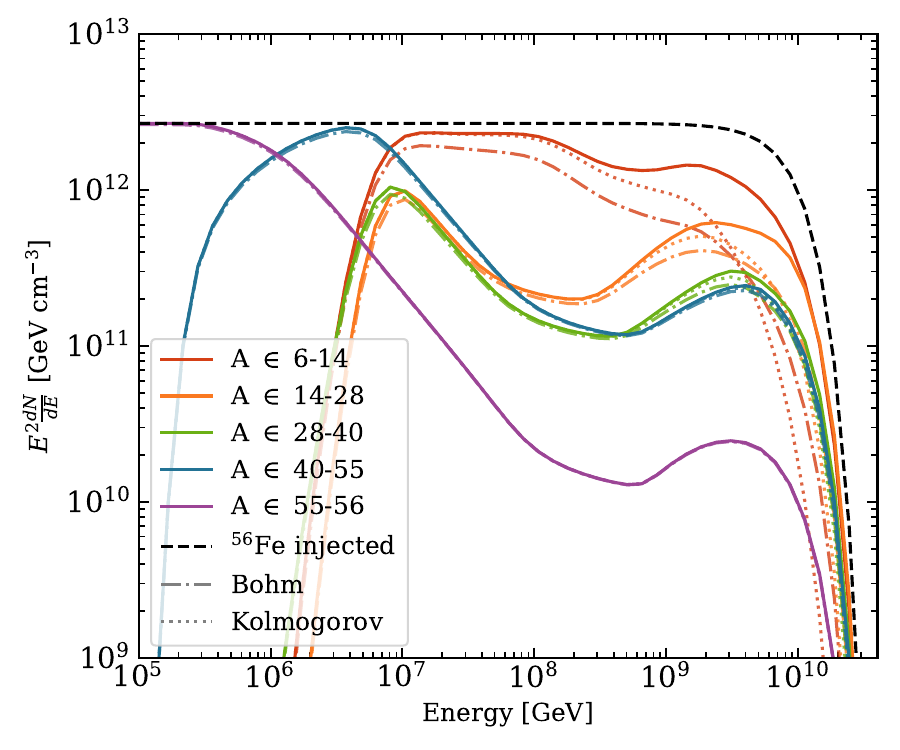}
    \caption{UHECR spectral densities of a GRB source in the optically thick scenario. Shaded regions and line styles indicate the effect of different model assumptions. Left: Influence of a time-varying injection with a fixed total injection. The case of a constant injection (solid lines) is contrasted with a quadratically increasing injection (dotted bound) and a linearly decreasing injection (dash-dotted bound). Right: Influence of rigidity-dependent escape assumptions. The solid lines represent advective escape, as in the left figure. The dash-dotted lines show the Bohmian diffusion, and the dotted lines show the effect of diffusion under a Kolmogorov-distributed turbulent magnetic field. Additional details are given in the text and in Appendix~\ref{sec:source_details}.}
    \label{fig:source_spectrum_escape}
\end{figure*}

\subsection{Reverse propagation}
\label{subsec:backpropagation}

Under certain conditions, the direct Markov jump process that describes nuclear cascades can be reversed. This is particularly relevant to the problem of inferring the composition of cosmic rays at their source, given the composition measured on Earth.

The simplest case of the reverse-propagation process is the quasi-stationary regime. In Markov jump processes, the stationary distribution, $\boldsymbol{\phi}^s$, is determined by the condition $\boldsymbol{\phi}^s\boldsymbol{\Lambda}(\gamma)=\boldsymbol{0}$, meaning that the composition remains unchanged as time evolves. However, in the cascades discussed here, all nuclear states are transient; therefore, no such stationary distribution exists. Nevertheless, a quasi-stationary state can be reached with a corresponding distribution $\boldsymbol{\tilde \phi}^s$, defined by the relation $\boldsymbol{\tilde\phi}^s\boldsymbol{\Lambda}(\gamma)=-\tilde\lambda^s \boldsymbol{\tilde\phi}^s$, which implies that reverse-propagation preserves the Markov property. The corresponding reverse interaction matrix can be easily constructed via
\begin{equation}
    \boldsymbol{\tilde \Lambda}_r(\gamma) = \mathrm{diag}(\boldsymbol{\tilde \phi}^s)^{-1} \boldsymbol{\Lambda}(\gamma)^T \mathrm{diag}(\boldsymbol{\tilde \phi}^s) .
\end{equation}
By construction, $\boldsymbol{\tilde \phi}^s$ is the same for both the forward and reverse processes. The reverse process can then be computed with $\boldsymbol{\tilde \Lambda}_r(\gamma)$ integrating Kolmogorov's differential equation or by building the probability distributions of distance until absorption as above. Here, absorption corresponds to probing the original species or composition assumption. 

Figure~\ref{fig:backpropagation} illustrates the likelihood of observing a cosmic ray nucleus of a given energy from different distances under different assumptions about the original species. These likelihoods were computed using Kolmogorov's differential equation to evolve the probability vector. The likelihood for each species is the point probability for that species as a function of distance, normalized to a common value for comparison with the other species. However, the relative likelihoods can also be inferred by a different approach. As expected, the heavier the assumed original species, the greater the maximum likelihood distance. This approach can be used to estimate the origin of individual events with extreme energies~\citep{Morejon:2024rtq}, such as the recent Amaterasu particle detected by the Telescope Array~\citep{doi:10.1126/science.abo5095}. 

Assuming a quasi-stationary distribution is a very specific condition that may not be met in reality. Verifying this assumption for the observed UHECR spectrum would require a level of precision in energy and composition that is currently out of experimental reach. A more general approach is to numerically solve Kolmogorov's differential equation for the inverse process.

\subsection{UHECR sources}
\label{subsec:sources}

There are two ways to apply the approach of solving Kolmogorov's differential equation to model UHECR sources. The simplest method is to compute the distributions until absorption to e.g.~determine the probability of escape of a given species~\citep{Morejon:20239X,Morejon:2025nb}. For example, the probability vector $\boldsymbol{\phi}_{\rm esc}$ with the composition at the time of escape can be obtained applying Eq.~\ref{eq:KolmogorovEqs} on an assumed injected composition, $\boldsymbol{\phi}_{\rm inj}$,
\begin{equation}
    \boldsymbol{\phi}_{\rm esc}(\gamma) = \boldsymbol{\phi}_{\rm inj} \boldsymbol{P}^{t_s}(\gamma)
\end{equation}
with  $t_s\approx L_s/c$ as the characteristic crossing time of the source. Here, the rates contained in $\boldsymbol{\Lambda}$ (and therefore $\boldsymbol{G}(\gamma)$, used to find $\boldsymbol{P}(\gamma)$) are computed with the source target photons (e.g.~a broken power law). The effects of CI and DI can be taken into account (as discussed in Sect.~\ref{sec:continuous_losses}) to define the boost evolution and the corresponding evolution of $\boldsymbol{G}(\gamma(t))$. Furthermore, additional effects can be taken into consideration, such as the impact of different assumptions for the escape. For instance, if the distribution of trajectory lengths until escape, $F_{\rm esc}(\gamma, L)$, is known (a cumulative density as a function of trajectory lengths and the boost), the escape probability vector as a function of the boost would be
\begin{equation}
    \boldsymbol{\phi}_{\rm esc}(\gamma) = \boldsymbol{\phi}_{\rm inj} \int_0^L\boldsymbol{P}(\gamma)^{ L'/c} \left(1 - F_{\rm esc}(\gamma, L')\right) dL' .
\end{equation}

This expression assumes that changes in rigidity during successive disintegrations can be ignored, but this needs to be assessed for each specific scenario. When this is not the case, a more nuanced approach is also available (as illustrated in Sect.~\ref{subsec:magnetic_deflections} for propagation). 

The second method, of special interest, is simulating the time evolution of the composition in sources with a time-dependent cosmic ray injection. This type of modeling has been achieved with full nuclear cascades (e.g., NEUCOSMA~\citep{Biehl2018CosmicCascade,Rodrigues2017}) by numerically integrating Eq.~\ref{eq:matrix_SODE}, yielding time-dependent spectral densities for each nuclear species. This task can also be accomplished using our stochastic approach with regularization; namely, on the condition that all jumps occur at regular intervals of the elapsed time or distance, which can be arbitrarily small. Such assumption is valid given the large luminosities, which justify a continuous limit approach. This allows us to treat the occupation probabilities as volumes in a system of equations such as Eq.~\ref{eq:matrix_SODE} describing a fluid, where the changes in the occupation probability represent the amounts transferred between species as a function of time. In these cases, $\boldsymbol{\tilde Q}^{\rm ext}(\gamma, t)$ represents the injection vector, which may be a function of time, and is typically a power-law of the energy or boost. The time evolution of the probability vector at a later time, $t'$, is given as above $\boldsymbol{\tilde Q}^{\rm ext}(\gamma, t)\boldsymbol{P}^{t'-t}(\gamma)$, and the total yield can be computed by integrating over a certain injection time, $t_{\rm inj}$, or by taking a convolution product, 
\begin{equation}
    \boldsymbol{N}(\gamma, t_{\rm inj}) = \int_0^{t_{\rm inj}} \boldsymbol{\tilde Q}^{\rm ext}(\gamma, t')\boldsymbol{P}^{t_{\rm inj} - t'}(\gamma) dt' ,
\end{equation}
where $\boldsymbol{N}(\gamma, t_{\rm inj})$ is a vector with the final yields for each species in the cascade as a function of the boost. In simple cases where the DIs are negligible, we have $\boldsymbol{P}^t = e^{\boldsymbol{G}t}$, as discussed previously. However, the general form, including the DIs, requires computing $\boldsymbol{P}^t$ numerically or following a similar approach, as described in Sect.~\ref{subsec:dispersive_inhomogeneities}. This expression allows for arbitrary choices in the temporal evolution of the injection.

Figure~\ref{fig:source_spectrum_escape} illustrates the densities for different mass groups obtained by modeling a GRB example in the optically thick scenario \citep{Biehl2018CosmicCascade}. More details are given in Appendix~\ref{sec:source_details}. In this case, the injection rate vector $\boldsymbol{\tilde Q}^{\rm ext}(\gamma, t)$ consists of only one species, $^{56}$Fe, having a power-law dependence on the boost with a cutoff, while its norm $C'$ is determined by energy arguments. The effect of the temporal behavior is illustrated in Fig.~\ref{fig:source_spectrum_escape} (left), using a constant injection of cosmic rays as the baseline (solid lines), a quadratically increasing injection as the lower limit (dotted bound of the shaded regions), and a linearly decreasing injection (dash-dotted bound of the shaded regions). All parameters were fixed by requiring the same total injection over the fixed injection time $t_{\rm inj}$. This example assumes that nuclei escape after propagating a characteristic distance (or time scale), corresponding to an advective escape.

In addition, as discussed above, other assumptions for the escape can be included. Figure~\ref{fig:source_spectrum_escape} (right) presents the effect of different escape assumptions computed according to
\begin{equation}
    \boldsymbol{N}(\gamma, t_{\rm inj}) = \boldsymbol{\tilde Q}^{\rm ext}(\gamma) \int_0^{t_{\rm inj}} \boldsymbol{P}^{t_{\rm inj} - t'}(\gamma) \circ \left( 1 - \boldsymbol{F}_{\rm esc}(R, t') \right)dt' ,
\end{equation}
where the injection rate used corresponds to the constant injection of iron, as shown in the left plot. The matrix $\boldsymbol{F}_{\rm esc}(R, t')$  describes the probability distribution of escape as a function of rigidity $R = E/Z = \gamma/\kappa$, which varies with the nuclear species as $\kappa=Z/A$. The operation $\circ$ denotes the element-wise product of the two matrices, each evaluated at the time since injection $t'$. The solid lines represent the advective escape, as shown in Fig.~\ref{fig:source_spectrum_escape} (right). Two other line styles represent alternative assumptions of rigidity-dependent escape: a Bohmian diffusion case and diffusive escape under a Kolmogorov-distributed turbulent magnetic field. In both cases, escape is exponentially distributed according to $F_{\rm esc}=1 - \exp(-t/t_{\rm diff})$ with dependencies $t_{\rm diff} = 3\cdot10^6 / R$ for the Bohmian case (where the diffusion coefficient is proportional to rigidity) and $t_{\rm diff} = 2\cdot10^2 / R^{1/3}$ for the Kolmogorov case (where the diffusion coefficient is proportional to the cube root of rigidity).

One advantage of this approach is that it allows for consistent handling of interactions within the source and during propagation within a single model. This differs from current approaches, which simulate each environment separately, and allows us to fit the spectrum and composition of cosmic rays by including the source parameters in the minimization, such as the source's optical thickness or the injected composition. For example, this approach enables us to connect UHECR emissions to nuclear cascade models, which describe optical observations of kilonovae resulting from neutron star mergers and associated GRBs.

\subsection{Magnetic deflections and distribution of arrival direction}
\label{subsec:magnetic_deflections}

The probabilistic disintegration of nuclei during propagation also affects the arrival directions of cosmic rays. Previous studies \citep[e.g.][]{Lee1995,Waxman1996} have discussed the angular deviations that occur during UHECR propagation under the influence of extragalactic magnetic fields (EGMFs) in the small-angle scattering regime (i.e., when the gyroradius exceeds the EGMF coherence length $\lambda_B$). However, these and similar studies neglect energy losses and disintegrations by assuming a constant rigidity $R$, which leads to the mean squared angular deviation formula $\langle\Delta \theta \rangle^2 \approx \frac{4}{\pi^2}\frac{B^2}{R^2}\lambda_Bd$~\citep{Lee1995}, where $d$ denotes the distance to the source and $B$ the strength of the magnetic field. 

This expression is sometimes used even in the presence of disintegrations, arguing that changes in nuclear species do not affect the rigidity, since $R = E/Z = \gamma/\kappa$, $\gamma$ is conserved and $\kappa=Z/A$ can be considered 0.5 for most stable species up to iron. However, the actual variation of $\kappa$ in nuclear cascades can be up to 30~\% of the mean value, and the changes are stochastic as the cascade itself. Currently, there is no expression for the angular deviation that incorporates disintegrations. The most realistic treatment involves Monte Carlo simulations with a code that includes magnetic effects and disintegrations, such as CRPropa \citep{2022JCAP...09..035A}.

\begin{figure}
    \centering
    \includegraphics[width=0.45\textwidth]{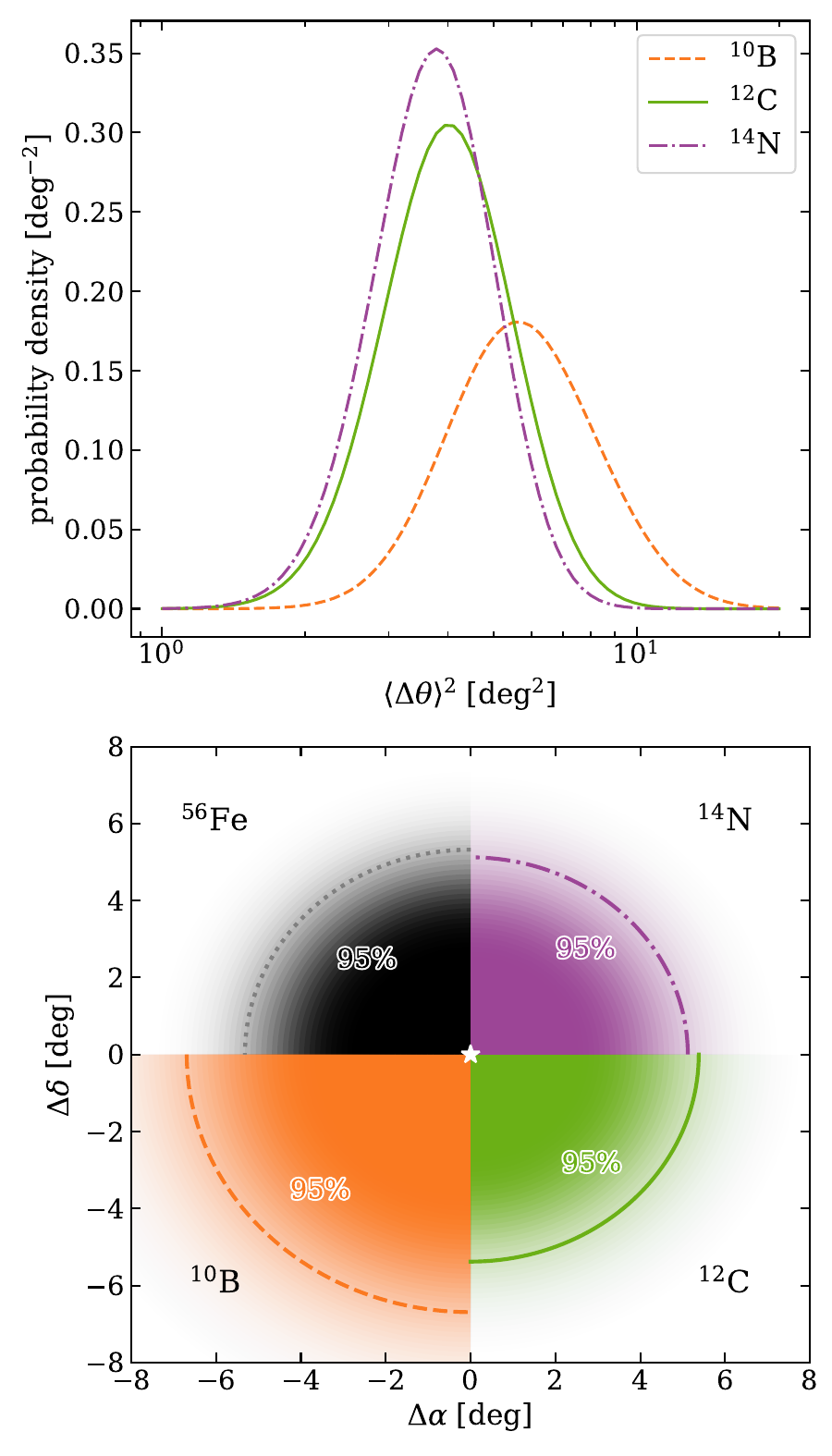}
    \caption{Impact of disintegration on the dispersion angle of different arriving nuclei resulting from the disintegration of 300\,EeV $^{56}$Fe nuclei over a distance of 50\,Mpc. Top: Distributions of mean squared deflection angles for different secondaries with similar mass and charge to that of $^{12}$C. Bottom: The angular distributions of the same products in galactic coordinates are compared to the distribution for the parent $^{56}$Fe disregarding disintegration. The star represents the position of the source in the sky and the circular lines denote the 95\% confidence limit.}
    \label{fig:angular_distance_likelihoods}
\end{figure}

Using the stochastic description in this work, we can derive an analytic expression for the expected angular spread of individual products. The mean squared angular deviation for each species $i$ in the cascade is given by $\theta_{\rm ms, i}=\langle\Delta \theta_i \rangle^2 \approx \frac{4}{\pi^2}\frac{B^2\kappa_i^2}{\gamma^2}\lambda_Bd_i$, where $d_i$ is the rectilinear distance that species $i$ travels from creation until interacting. The distribution of the total mean squared angular deviation is the sum of all the stochastic deviations experienced by each species $\theta_{\rm ms} = \sum_i\theta_{\rm ms, i}$ (i.e., it is a stochastic variable itself). This also implies that $\theta_{\rm ms}$ is a phase-type distribution which can be obtained by applying a "transformation via rewards" to the distribution of the distance traveled until absorption. In this transformation, the "reward" is the deviation per unit distance for each species, which is given by $\frac{4}{\pi^2}\frac{B^2\kappa_i^2}{\gamma^2}\lambda_B$. It should be noted the linear dependence between the propagation distance and $\theta_{\rm ms, i}$, which is a requirement of this type of transformations. Examples of the obtained distributions for $\theta_{\rm ms}$ are shown in Fig.~\ref{fig:angular_distance_likelihoods} (top) for the total mean squared angular deviation accumulated until reaching $^{10}$B, $^{12}$C and $^{14}$N in the propagation of a 300~EeV $^{56}$Fe nucleus over a propagation distance of 50~Mpc. The products were chosen as species with similar masses and charges are produced at comparable distance scales, although their production in this example ranges from 10-100~Mpc, hence the choice of 50~Mpc for $^{56}$Fe.

Without disintegration (i.e., constant rigidity), the distribution of the angular deviation $\theta$ from the source direction $P(\theta~|~\theta_{\rm ms})$ is a Rayleigh distribution with parameter $\theta_{\rm ms} =\langle\Delta \theta \rangle^2$. The inclusion of disintegration is reflected in the probability distribution of mean squared deviations $f(\theta_{\rm ms})$, and the sought distribution of the angular deviation is a mixture of Rayleigh distributions $P(\theta)=\int P(\theta~|~\theta_{\rm ms})f(\theta_{\rm ms})d\theta_{\rm ms}$, using the density $f(\theta_{\rm ms})$ obtained with a transformation via rewards, 
\begin{equation}
    P(\theta) = \int_0^{\infty} 2\frac{\theta}{\theta_{\rm ms}} e^{ -\frac{\theta^2}{\theta_{\rm ms}}} f(\theta_{\rm ms}) d\theta_{\rm ms} .
\end{equation}
The bottom plot in Fig.~\ref{fig:angular_distance_likelihoods} shows the distributions of angular deviation from the source's direction for each of the aforementioned products. For comparison, the distribution of the initial iron after 50~Mpc neglecting disintegrations (as is typically assumed) as the initial iron nuclei do not survive beyond a few megaparsecs. The differences in angular distributions are quantified by the 95\% containment angle, $\theta_{95}$, represented with the corresponding lines matching the top plot. Differences between secondaries reflect mainly the differences in the range of production distances, as lighter products tend to have  larger means of the distance distributions. The angular deviation for $^{14}$N is slightly reduced as it is produced at a mean distance of 42~Mpc, whereas for $^{12}$C the value is 46~Mpc and in the case of $^{10}$B is much larger at 68~Mpc and spans beyond 100~Mpc. An important consequence of this result is that precisely identifying the observed species could drastically alter their association with existing astrophysical objects as possible origins.

It should be noted that studying these types of distributions using Monte Carlo methods is extremely computationally expensive. This is because the phase space of distances, starting nuclei, and final products would require multiplying the already large number of candidates to be simulated by a factor of about 1000 to obtain an adequate description of these cascades (see Appendix~\ref{sec:efficiency}).

\section{Conclusions}
\label{sec:conclusions}

Until now, the stochasticity of UHECR interactions has been addressed using Monte Carlo approaches, which are limited by the availability of appropriate computational resources. This work demonstrates that interactions of UHECRs with photon fields in astrophysical scenarios can be described analytically with arbitrary precision. This description has additional advantages, including the ability to obtain closed-form probability distributions, such as the distance until loss of a number of nucleons and the deflections of UHECRs in the EGMF including interactions and secondary nuclei.

The stochastic approach presented here provides physical insights; for example, the regularity of the distributions under variations of the boost, or the mass dependence of the full disintegration distances, which can span from logarithmic to linear depending on the number of available disintegration channels. These insights are possible because of the closed form of the moments of the distributions and the distributions themselves. Furthermore, we have established a clear distinction of the types of cascades produced by the different types of nuclear cross-section models used in UHECR astrophysics. The classification of the cascades types relates to the number of disintegration channels and determines the type of distributions applicable and the impact on cetrain relations, such as the disintegration evolution with  distance. Thus, this approach can enhance our understanding of the connections between the cross-section data and the propagation of UHECRs.

The detailed knowledge of the disintegration cascades strengthens the methods used to connect observations to the possible origins. We have implemented precise expressions for horizons, which can help establish distances of likely origin using priors on the original composition. The distributions of angular deviation derived here can discriminate between nuclei emitted by the source and their disintegration products. In addition, we show how the precise description of the cascade allows us to construct backpropagation processes to explore the connections of observed nuclei to their possible parent species.

Our approach is not limited to studying probabilities for compound quantities; it can also be applied to interactions in scenarios where stochasticity is less important due to the large number of expected events. For example, we computed the cosmic ray production for a GRB source scenario and demonstrated how source properties such as the magnetic field and time-dependent injection affect the variability of the produced spectrum. Furthermore, this approach could enable a combined source-propagation framework that circumvents ``coupling defects'' between the two scenarios, given that propagated compositions are currently often reduced to a few mass groups. At the same time, a common treatment of both the in-source and propagation processes allows for consistency in the cross-section and decay tables without the need for the simplifications that are often employed in existing frameworks.

The expressions presented here (along with additional utility functions) have been made available through an open-source Python package called Cosmic Ray Stochastic Interactions for Propagation (CRISP). The package is meant to provide a common implementation for the community to facilitate the computation of quantities of interest in UHECR astrophysics and facilitate the reproducibility of the results shown in this work.  Future research will be focused on studying the observed UHECR spectrum and composition with fewer assumptions (e.g., excluding the spectral index or source evolution) and exploring the sensitivity of the fitted parameters.

\begin{acknowledgements}

This work has received funding via the grant MultI-messenger probe of Cosmic Ray Origins (MICRO) from the DFG through project number 445990517. Further support was provided by Institut Pascal at Université Paris-Saclay within the program “Investissements d’avenir” ANR-11-IDEX-0003-01, the P2I axis of the Graduate School of Physics of Université Paris-Saclay, as well as IJCLab, CEA, IAS, OSUPS, and APPEC.

\end{acknowledgements}

\vspace{5mm}
\software{
This work employed the software packages Astropy~\citep{astropy:2013,astropy:2018,astropy:2022}, Matplotlib~\citep{Hunter:2007}, Numpy~\citep{harris2020array}, Scipy~\citep{2020SciPy-NMeth}.
}

\bibliography{references}
\bibliographystyle{aasjournal}

\appendix

\section{Derivation of the canonical form}
\label{sec:deriving_canonical}

The canonical form corresponding to an RSeC describes the probability that a nucleus of mass number $A$ will interact $k$ times over a trajectory length $L$. In these cascades, each interaction results in the loss of one nucleon, reducing the mass number by one. Thus, after the final interaction, the remaining nucleus has $A-k$ nucleons. Additionally, the interaction rates for all nuclei satisfy the regularity condition (i.e., the relations $\lambda_{A_i} = \frac{A_j}{A_i} \lambda_{A_j}=A_j \cdot \lambda_1$ hold) with $\lambda_1$ the interaction rate per nucleon.

The probability that nucleus $A$ will interact within the differential length $dx$ at position $x$ from a starting point follows an exponential distribution,
\begin{equation}
    f_{A \to A-1} (x) = \lambda_A e^{-\lambda_A x},
\end{equation}
with $\lambda_A$ the interaction rate per unit length. The probability $P_{A \to A-1} (x \leq L)$ that nucleus $A$ will interact within a trajectory length smaller than or equal to $L$ is given by the corresponding distribution function,
\begin{equation}
    F_{A \to A-1} (L) = 1 - e^{-\lambda_A L} .
\end{equation}
To describe the probability that the sequence of nuclei $\{A, A-1, ..., A-k+1\}$ occurs within the path length $L$, we must integrate over all possible intermediate trajectory lengths corresponding to the distances covered by each nucleus until interacting $\{x_A, x_{A-1}, ..., x_{A-k+1}\}$, such that $L = x_A + x_{A-1} + ... + x_{A-k+1}$. In essence, this involves finding the distribution describing $L$ as the sum of $k$ exponentially distributed functions with rate parameters $\{\lambda_A, \lambda_{A-1}, ..., \lambda_{A-k+1}\}$, given by the integral,
\begin{equation}
    \begin{split}
        F_{A \to A-k}(L) = \int_0^{L} dx_A f_{A \to A-1} (x_A) \int_0^{L-x_A} dx_{A-1} f_{A-1 \to A-2} (x_{A-1}) \int_0^{L-x_A-x_{A-1}} dx_{A-2} \int ... \\ \int_0^{L - \sum_{l=0}^{k-1}x_{A-l}} dx_{A-k+1} f_{A-k+1 \to A - k} (x_{A-k+1}) .
    \end{split}
\end{equation}
The expressions for the first few values of $k$ yield
\begin{align}
     F_{A \to A-1} (L) =& 1 - e^{-\lambda_A L},\\
     F_{A \to A-2} (L) =& 1 + (A-1)e^{-\lambda_A L} - Ae^{-\lambda_{A-1} L},\\
     F_{A \to A-3} (L) =& 1 - \frac{(A-1)(A-2)}{2}e^{-\lambda_A L} + A(A-2)e^{-\lambda_{A-1} L} - \frac{(A-1)(A-2)}{2}e^{-\lambda_{A-2}L},\\
     F_{A \to A-4} (L) =& 1 + \frac{(A-1)(A-2)(A-3)}{6}e^{-\lambda_A L} - \frac{A(A-2)(A-3)}{2}e^{-\lambda_{A-1} L}, \\ 
     & -\frac{A(A-1)(A-3)}{2}e^{-\lambda_{A-2}L} +\frac{A(A-1)(A-2)}{6}e^{-\lambda_{A-3}L},\\
     F_{A \to A-5} (L) =& ...\\
\end{align}
This can be generalized, by induction, in the form
\begin{equation}
    \begin{split}
        F_{A \to A-k} (L) =  1 + k {A \choose k} \sum_{l=0}^{k-1} (-1)^{k-l} {k-1 \choose l} \frac{e^{-\lambda_{A-l}L}}{A-l} .\\
    \end{split}
\end{equation}
The density function can be found by deriving with respect to $L$,
\begin{equation}
    f_{A \to A-k} (L)=\frac{d}{dL}F_{A \to A-k} (L) = -k \lambda_1 {A \choose k} e^{-\lambda_AL}\sum_{l=0}^{k-1} (-1)^{k-l} {k-1 \choose l} e^{\lambda_l L},
\end{equation}
where the binomial theorem has been employed. Rearranging some terms and employing some known identities leads to Eq.~\ref{eq:kan_density}. 
Noting that $k {A \choose k} = (A - k + 1) {A \choose k-1}$ and rewriting $e^{-\lambda_A L}$ as $e^{-\lambda_{A-k+1} L}e^{-\lambda_{k-1} L}$ brings the second factor into the sum. The expression then becomes
\begin{equation}
    f_{A \to A-k} (L)=\lambda_{A-k+1} {A \choose k-1} e^{-\lambda_{A-k+1} L}\sum_{l=0}^{k-1} (-1)^{k-1-l} {k-1 \choose l} e^{-\lambda_{k-1-l} L},
\end{equation}
where it is apparent from the binomial identity that the sum is equivalent to $(1-e^{-\lambda_1L})^{k-1}$, as in Eq.~\ref{eq:kan_density}.

Other forms of this expression can be found in terms of known functions with precomputed values. Setting the variable $\xi = e^{-\lambda_1 L}$ we have $\frac{d}{dL} = -\lambda_1 \xi \frac{d}{d\xi}$. Some terms change as $e^{-\lambda_A L} = \xi^A$ and $\left(e^{\lambda_l L} - 1\right) = \left(1 - e^{-\lambda_l L}\right)e^{\lambda_l L}$, resulting in
\begin{equation}
    f_{A \to A-k} (\xi) = - (A - k + 1) {A \choose k-1} \xi^{A - k} \left(1 - \xi\right)^{k-1},
\end{equation}
where the relation $(A -k+1){A \choose k-1} = 1/\mathrm{B}_{\xi}(A-k+1, k)$ was used, where $\mathrm{B}(\alpha, \beta)$ is the beta function. This form corresponds to the beta distribution $\mathcal{B}(\alpha, \beta)$, namely, computing the RSeC distribution requires only evaluating $\mathcal{B}(\alpha, \beta)$ with the appropriate arguments.

The corresponding distribution function,
\begin{equation}
    F_{A \to A-k} (\xi) = 1 - \int_0^{\xi} \frac{{\xi'}^{A-k} (1 - \xi')^{k-1}}{\mathrm{B}(A-k+1, k)} d\xi' = 1 - \mathrm{B}_{\xi}(A-k+1, k) = \mathrm{B}_{1-\xi}(k, A-k+1),
\end{equation}
is given in terms of the incomplete beta function, $\mathrm{B}_z(\alpha, \beta)$, which has known expressions for the expected value $\mathbb{E}[\xi]=\frac{\alpha}{\alpha+\beta}$ and the variance $\mathrm{Var} [\xi]=\frac{\alpha \beta}{(\alpha +\beta)^2 (\alpha+\beta+1)}$. Substituting $\xi$ and the appropriate values for $\alpha$ and $\beta$ in these expressions yields 
 \begin{equation}
     \mathbb{E}[\lambda_1 L]_{A \to A-k} = \ln{\left( \frac{A+1 }{A-k+1}\right)},
 \end{equation}
\begin{equation}
    \mathrm{Var} [\lambda_1L]_{A \to A-k} = -\ln \left(1 -  \frac{k(A-k+1)}{(A+1)^2(A+2)} \right) .
\end{equation}
The distribution $\mathcal{B}(\alpha, \beta)$ and the binomial distribution are well-known and interconnected. They help us interpret functions in terms of the superposition model where the interaction probability of an individual nucleon determines the probability that the nucleus suffers a given number of interactions.

\section{Full disintegration distance for ISeCs and CoCs} 
\label{sec:expected_distance}

The expected distance until full disintegration for RSeCs involves the logarithm of the initial mass, as seen in Appendix~\ref{sec:deriving_canonical}. This expression can be compared to the one for ISeCs, by transforming the expression of the expected distance for the latter,
\begin{equation}
     \mathbb{E}[\lambda_1L] = \lambda_1 \sum_{k=1}^{A} \frac{1}{k\lambda_1} + \left(\frac{1}{\lambda_k} - \frac{1}{k\lambda_1}\right)
     = \sum_{k=1}^{A} \frac{1}{k} + \left(\frac{1}{\tilde k} - \frac{1}{k}\right)
     = H_A + \sum_{k=1}^{A} \frac{k - \tilde k}{k\tilde k} ,
\end{equation}
where $H_A$ is the harmonic number and the second term contains the deviation from the regular case. Examining this sum closely, we see that the deviations $\chi_k=\frac{k - \tilde k}{\tilde k}$ are expected to be randomly distributed with a null mean and modules smaller than unity. We may place a limit on the second term by taking the largest deviation, $\chi^{\rm max}$, and factoring it out from the sum. It is clear that all ratios satisfy $\left|\chi_k/\chi^{\rm max} \right| \leq 1$. Thus, if we replace all ratios by unity with the corresponding sign, we can assert that the module of this sum is strictly larger
\begin{equation}
    \left|\chi^{\rm max}\sum_{k=1}^{A} \frac{\chi_k/\chi^{\rm max}}{k}\right| < \left|\chi^{\rm max}\sum_{k=1}^{A} \frac{\varepsilon_k}{k}\right| < \left|\chi^{\rm max}\right| \left|\sum_{k=1}^{\infty} \frac{\varepsilon_k}{k} \right|,
\end{equation}
where $\left|\sum_{k=1}^{\infty} \frac{\varepsilon_k}{k} \right| \lesssim 3$ is the random harmonic series \citep{Schmuland01052003}, which is bounded. Therefore, we can conclude that for ISeCs $\mathbb{E}[\lambda_1L]$ differs from the behavior of RSeCs by no more than $3\left|\chi^{\rm max}\right|$.

The general expression, applicable to arbitrary CoCs, can be derived explicitly from the general expression (see Table~\ref{tab:moments}), $\mathbb{E}[\lambda_1L]=-\lambda_1\boldsymbol{\phi} \boldsymbol{\Lambda}^{-1}\boldsymbol{1}$, where $\boldsymbol{\phi}$ would have only injection of the heaviest nucleus. Hence, all entries are null except the heaviest species (first row), where it is 1. In such a case, the expression yields the sum of all the elements in the first row of $\boldsymbol{\Lambda}^{-1}$, which can be determined by solving for the first row $\boldsymbol{\Lambda}^{\rm T}\boldsymbol{x}=\boldsymbol{e}_1$ (where $\boldsymbol{e}_1$ is a vector with all entries null except the first one being 1) and where each entry can be solved iteratively, yielding
\begin{align}
        x_1 &= \frac{1}{\lambda_{S_1}},\\
        x_2 &= \frac{\lambda_{S_1 \to S_2}}{\lambda_{S_1}\lambda_{S_2}},\\
        x_3 &= \frac{\lambda_{S_1 \to S_3}}{\lambda_{S_1}\lambda_{S_3}} + \frac{\lambda_{S_1 \to S_2}\lambda_{S_2 \to S_3}}{\lambda_{S_1}\lambda_{S_2}\lambda_{S_3}},\\
        x_4 &= \frac{\lambda_{S_1 \to S_4}}{\lambda_{S_1}\lambda_{S_4}} + \frac{\lambda_{S_1 \to S_2}\lambda_{S_2 \to S_4}}{\lambda_{S_1}\lambda_{S_2}\lambda_{S_4}} + \frac{\lambda_{S_1 \to S_3}\lambda_{S_3 \to S_4}}{\lambda_{S_1}\lambda_{S_3}\lambda_{S_4}} + \frac{\lambda_{S_1 \to S_2}\lambda_{S_2 \to S_3}\lambda_{S_3 \to S_4}}{\lambda_{S_1}\lambda_{S_2}\lambda_{S_3}\lambda_{S_4}},\\
        x_5 &= \frac{\lambda_{S_1 \to S_5}}{\lambda_{S_1}\lambda_{S_5}} + \frac{\lambda_{S_1 \to S_2}\lambda_{S_2 \to S_5}}{\lambda_{S_1}\lambda_{S_2}\lambda_{S_5}} + \frac{\lambda_{S_1 \to S_3}\lambda_{S_3 \to S_5}}{\lambda_{S_1}\lambda_{S_3}\lambda_{S_5}} + \frac{\lambda_{S_1 \to S_4}\lambda_{S_4 \to S_5}}{\lambda_{S_1}\lambda_{S_4}\lambda_{S_5}} + ... \\
        & \frac{\lambda_{S_1 \to S_2}\lambda_{S_2 \to S_3}\lambda_{S_3 \to S_5}}{\lambda_{S_1}\lambda_{S_2}\lambda_{S_3}\lambda_{S_5}} + \frac{\lambda_{S_1 \to S_2}\lambda_{S_2 \to S_4}\lambda_{S_4 \to S_5}}{\lambda_{S_1}\lambda_{S_2}\lambda_{S_4}\lambda_{S_5}} + \frac{\lambda_{S_1 \to S_3}\lambda_{S_3 \to S_4}\lambda_{S_4 \to S_5}}{\lambda_{S_1}\lambda_{S_3}\lambda_{S_4}\lambda_{S_5}} + \frac{\lambda_{S_1 \to S_2}\lambda_{S_2 \to S_3}\lambda_{S_3 \to S_4}\lambda_{S_4 \to S_5}}{\lambda_{S_1}\lambda_{S_2}\lambda_{S_3}\lambda_{S_4}\lambda_{S_5}},\\
        x_6 &= ...
\end{align}
By generalizing and adding the above expressions, we obtain the formula for the expectation of distance until full disintegration,
\begin{equation}
        \mathbb{E}[\lambda_1L]
        = -\lambda_1\boldsymbol{\phi} \boldsymbol{\Lambda}^{-1}\boldsymbol{1}
        =\frac{\lambda_1}{\lambda_{S_1}} \sum_{n=1}^{|\{S\}|} \sum_{\{\alpha, \beta, ...\} \subseteq {[n] \choose k}}^{n \choose k} \frac{\lambda_{S_1 \to S_{\alpha}} \lambda_{S_{\alpha} \to S_{\beta}} ...}{\lambda_{S_{\alpha}}\lambda_{S_{\beta}} ...},
\end{equation}
where $[n] \choose k$ denotes the set formed by all $n \choose k$ combinations of $k$ indices chosen out of $n$ possible indices. This expression yields the ISeC result when there is only one nucleon emission channel, since only the term  ,
\begin{equation}
    \frac{\lambda_{S_1 \to S_2} \lambda_{S_2 \to S_3}...\lambda_{S_{n-1} \to S_n}}{\lambda_{S_1}\lambda_{S_2} ...\lambda_{S_n}} = \frac{1}{\lambda_{S_n}},
\end{equation}
survives for each $n$, yielding the inverse rate for the $n$-th species. Thus, the sum of these $n$ terms yields the expected result. Furthermore, it is evident that with the regularity condition this expression produces the result for RSeCs.

However, for CoCs this expression produces shorter values, as can be seen in the following example. Assuming that all species can undergo one-nucleon loss with a branching ratio $\chi$, so that the two-nucleon loss channel has a branching ratio of $1-\chi$, we have the term,
\begin{equation}
    \frac{\lambda_{S_1 \to S_2} \lambda_{S_2 \to S_3}...\lambda_{S_{n-1} \to S_n}}{\lambda_{S_1}\lambda_{S_2} ...\lambda_{S_n}} = \frac{\chi^{n-1}}{\lambda_{S_n}},
\end{equation}
similar to the previous case for one-nucleon loss only, and we have $n-2$ terms, where only one two-nucleon loss rate appears,
\begin{equation}
    \frac{\lambda_{S_1 \to S_2} \lambda_{S_2 \to S_3}...\lambda_{S_k \to S_{k+2}}...\lambda_{S_{n-1} \to S_n}}{\lambda_{S_1}\lambda_{S_2} ...\lambda_k ...\lambda_{S_n}} = \frac{\chi^{n-2}(1-\chi)}{\lambda_{S_n}},
\end{equation}
and we have $n-3 \choose 2$ terms where two two-nucleon loss rates appear,
\begin{equation}
    \frac{\lambda_{S_1 \to S_2} \lambda_{S_2 \to S_3}...\lambda_{S_k \to S_{k+2}}...\lambda_{S_l \to S_{l+2}}...\lambda_{S_{n-1} \to S_n}}{\lambda_{S_1}\lambda_{S_2} ...\lambda_k ...\lambda_l ...\lambda_{S_n}} = \frac{\chi^{n-3}(1-\chi)}{\lambda_{S_n}^2},
\end{equation}
and so on. Grouping terms ending in the same species, $S_n$, yields
\begin{equation}
    \frac{1}{\lambda_{S_n}} \left(\sum_{q=0}^{n-1} {n-1-q \choose q} \chi^{n-1-q}(1-\chi)^q\right) < \frac{1}{\lambda_{S_n}},
\end{equation}
which is shorter than the corresponding inverse rate for the $n$-th species. Indeed, comparing the expression to $(\chi+ (1-\chi))^{n-1}$ makes it clear that the coefficients in the expression are smaller than the binomial coefficients. This occurs for all $n$ terms and explains why the expectation values for the full disintegration distance are reduced in CoCs compared to serial cascades.

\section{Decoherence lengths for dispersive inhomogeneities during propagation}
\label{sec:di_decoherence_lengths}

To estimate the impact of DIs caused by Bethe-Heitler pair production losses (BHL) during propagation over cosmological distances, we compare survival probability distributions, one including and one excluding BHL. This is reasonable because, in photodisintegration cascades, the shortest length scale is determined by the injected species with the largest mass, as its interaction rates are usually the highest. Additionally, DI effects are produced in chains of more than one species, so if the effect is small for  the largest species in a cascade, it should be sufficiently small for all others.

The survival probability follows an exponential distribution, which, for sufficiently short distances, has a homogeneous interaction rate because the redshift does not change, namely, 
\begin{equation}
    F^{\rm H}(\delta) = 1 - \exp{\left(-\lambda_{S_0}((1+z)^2\gamma_c)\delta \right)} .
\end{equation}
In the presence of DIs, it is described by an inhomogeneous exponential distribution
\begin{equation}
    F^{\rm IH}(\delta) = 1 - \exp{\left(-\int \lambda_{S_0}((1+z)^2\gamma_c)d\delta \right)} ,
\end{equation}
where changes in the rate with $\gamma_c$ need to be included, as $\gamma_c=\gamma_c(\delta)$  varies with the cosmological thickness according to Eq.~\ref{eq:comboost_evol}.

\begin{figure}
    \centering
    \includegraphics[width=0.9\linewidth]{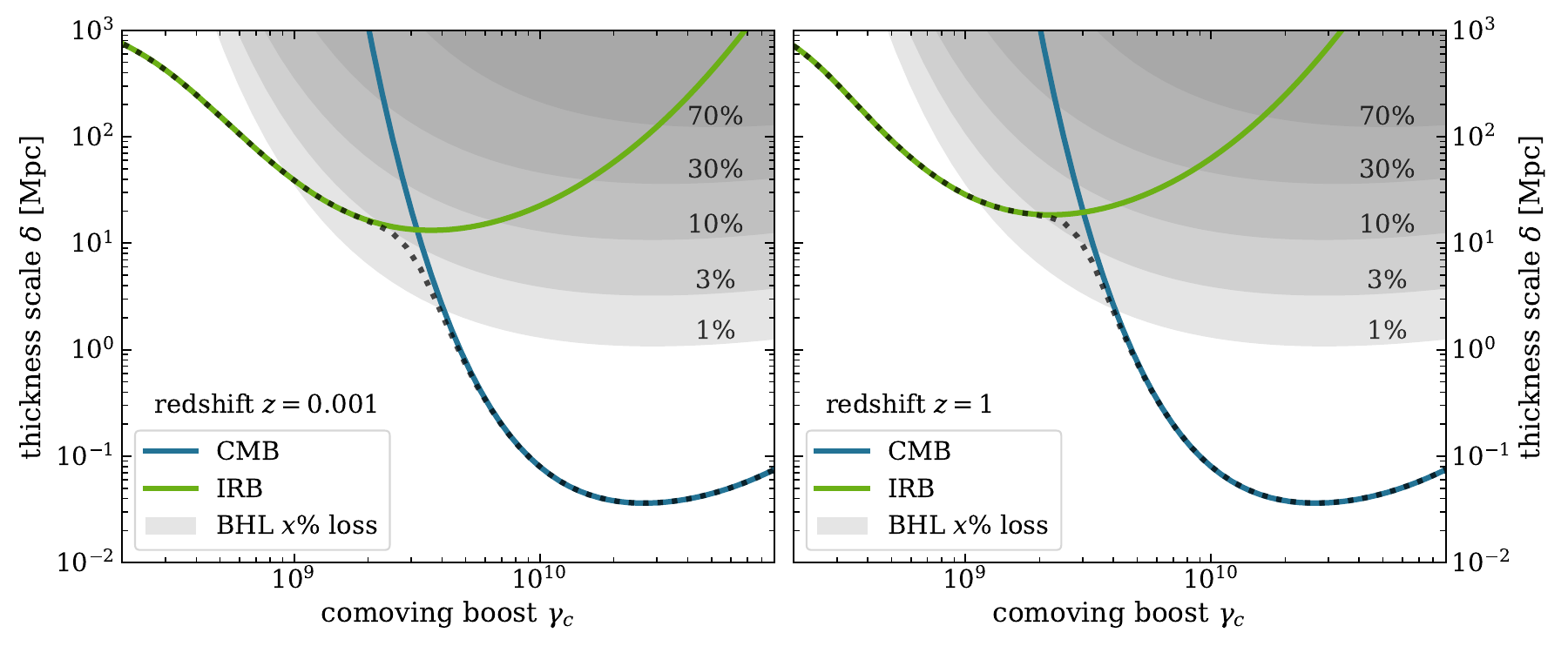}
    \caption{Thickness scales for iron to undergo different processes as a function of the comoving boost. Photodisintegration scales are represented by solid lines: in blue for interactions with the CMB and in green for interactions with the IRB. The thickness scales for the Bethe-Heitler pair production interactions are shown by shaded regions for certain values of relative energy energy loss in terms of percentage.}
    \label{fig:univrates_iron}
\end{figure}

Figure~\ref{fig:univrates_iron} shows the scales of cosmological thickness $\delta$ for different interactions experienced by an iron nucleus at two different redshifts. The total photodisintegration scale is shown by a black dotted line, which is the sum of the scales for interaction with the CMB (solid blue) and with the IRB (solid green, using the model by \cite{Gilmore_2012}). The scales are expressed in terms of the cosmological thickness; thus, the CMB interaction rates remain constant with redshift, while the IRB rates change according to $a(z)$ (see equations~\ref{eq:rates_on_z_g} and \ref{eq:cosmological_thickness}). The thickness scales for BHL are shown for multiple relative loss percentages. These scales do not change with redshift, as seen in Eq.~\ref{eq:comboost_evol}. The photodisintegration scales are shorter, implying that iron likely photodisintegrates before experiencing a relative energy loss of 3~\% due to BHL regardless of the Lorentz boost. As a consequence, photodisintegration rates practically do not change for the typical thickness scales required for this interaction. The same conclusion holds for a larger cosmological scales, however, at larger redshifts (e.g.~$z\geq2$) the photodisintegration distance scales for the IRB are comparable to the 10\% relative energy BHL scales. This stems from the changes in IRB density with redshift, as reflected in $a(z)$ (see Eq.~\ref{eq:rates_on_z_g}), while for the CMB the relation to BHL does not change. Nevertheless, to understand the impact a relative energy loss has on the probability distribution, we need to quantify the change in the interaction rate and its corresponding effect on the distribution.

In the first approximation of the Taylor expansion around $\delta=0$, the evolution of the disintegration rate yields, assuming the redshift can be considered constant,
\begin{equation}
    \lambda((1+z)^2\gamma_c(\delta)) = \lambda_0 + \frac{d\lambda}{d\delta}\bigg\rvert_{\delta=0}\delta = \lambda_0 + (1+z)^2\frac{d\lambda}{d\gamma_c}\frac{d\gamma_c}{d\delta}\bigg\rvert_{\delta=0}\delta = \lambda_0 + (1+z)^2\frac{d\lambda}{d\gamma_c}\bigg\rvert_{\delta=0}\gamma_c^0\frac{Z^2}{A}\frac{\beta_0(\gamma_c^0)}{c}\delta,
\end{equation}
where the starting comoving boost is $\gamma_c^0$. By using this expression in $F^{\rm IH}$, we can evaluate the difference between the homogeneous and inhomogeneous descriptions,
\begin{equation}
    (F^{\rm H}-F^{\rm IH})(\delta) = e^{-\lambda_0\delta} \left( \exp{\left({(1+z)^2\frac{d\lambda(\gamma_c)}{d\gamma_c}\bigg\rvert_{\delta=0}\gamma_c^0\frac{Z^2}{A}\frac{\beta_0(\gamma_c^0)}{c}\frac{\delta^2}{2}} \right)}  - 1 \right),
\end{equation}
which is a monotonic function of the cosmological thickness traversed $\delta$. With this expression we can evaluate the relative error incurred when neglecting BHL for a given length scale.

\begin{figure}
    \centering
    \includegraphics[width=0.85\linewidth]{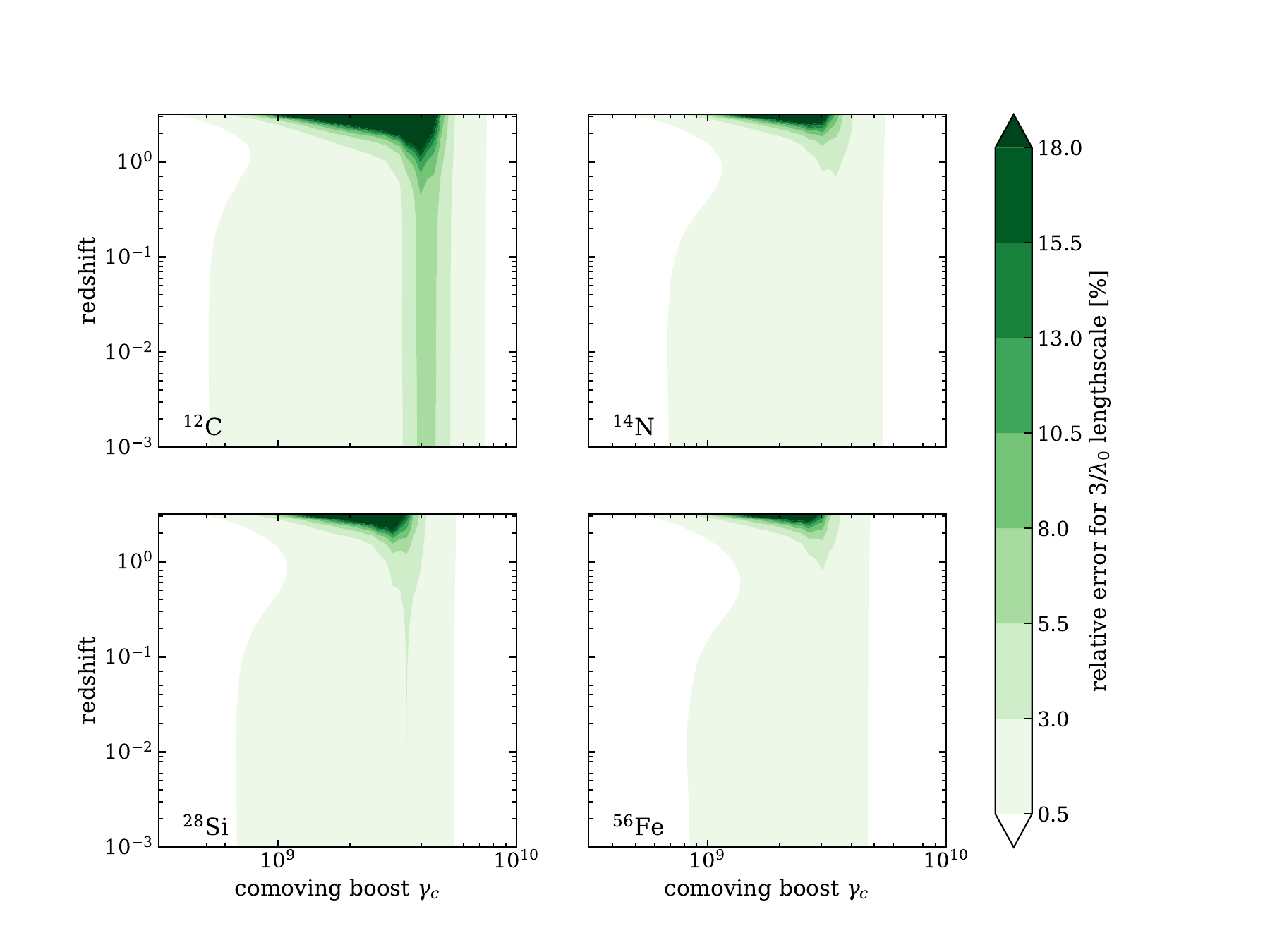}
    \caption{Percent error of neglecting DI for a length scale of three times the typical interaction thickness as function of the redshift and comoving boost.}
    \label{fig:opacity-limits}
\end{figure}

Figure~\ref{fig:opacity-limits} presents the relative error in neglecting the effect of BHL on a thickness scale comparable to three times the photodisintegration scale, which corresponds to a 95\% chance of interaction. Different nuclei are shown for comparison, and error bands are provided across a broad range of comoving boosts and redshifts. The error values vary across the phase space, but remain below 3\% for most of it, regardless of the nuclear species. The main differences concentrate around $\gamma_c \approx 3 - 4 \cdot 10^9$ since this is the region of transition from IRB to CMB interactions, where BHL can become important, as shown in figure~\ref{fig:univrates_iron}. These error values are acceptable for most applications. However, when greater precision is needed, the distributions can be computed numerically in the reduced region of the phase space where the impact is greater.

\section{Details of source examples}
\label{sec:source_details}

The source scenario considered in Fig.\,\ref{fig:source_spectrum_escape} is based on the GRB example in \citep{Morejon2019} and is discussed in more detail as the ``Optically Thick Case'' in \citep{Biehl2018CosmicCascade}. The GRB model is based on the ``fireball'' picture, in which cosmic rays are pre-accelerated and injected into the photon emission zone over a certain time interval, producing the emitted composition through photointeractions within this zone. The emission mechanism assumed is guided by the ``internal shock model'' in which relativistic shells collide, and a fraction of their kinetic energy powers the emission. The model conceives one such collision as a spherical shell expanding relativistically with a Lorentz factor $\Gamma = 300$, a radius of collision $R_c= 2\cdot 10^8$~km, and a volume $V_{\rm iso}=4\pi R_c^2 \Delta d'$, where $\Delta d'$ is the shell's thickness. The photon spectral number density is described by a broken power law between energies of 100~eV and 100~keV, with a power $-1$ below the break energy 1~keV and a power of $-2$ above the break. The energy density is $u_{\gamma}'=\frac{L_{\gamma}}{4\pi \Gamma^2 R_c^2}$, where $L_{\gamma}=10^{53}$~ergs/s is the luminosity. The magnetic field intensity is estimated assuming its energy density is comparable to the photon luminosity, $B' = \sqrt{\frac{2L_{\gamma}}{c \Gamma^2 R^2}}$. The injected cosmic ray species is $^{56}$Fe and the spectrum is given by $Q_{^{56}{\rm Fe}}'(E')=C'{E'}^{-2} \exp\left[-\left(E'/E'_{\rm max}\right)^2\right]$, where $E'_{\rm max}$ is the energy at which the acceleration rate is comparable to the sum of the rates for all energy losses, and $C'$ is determined by $\int_0^{10E'_{\rm max}}\tilde E Q_{^{56}Fe}'(\tilde E) d\tilde E=10u_{\gamma}' c / \Delta d'$ which stems from assuming that the baryonic loading factor (the ratio of the cosmic ray luminosity to the photon luminosity) is 10.

Additional considerations for the source models in this work involve the variability in the time dependence of the injection and the inclusion of rigidity-dependent escape rates. The baseline case for the effect of changes in the temporal form of the injection is a constant injection of cosmic rays of $C'(t)= C'_0 \approx 2.679 \cdot 10^{12}\, \rm{GeV\,cm^{-3}s^{-1}}$, and the variability is represented by the difference between the lower values of $C'(t)= 4(1 - (t-1)^2) \cdot 10^{12}\, \rm{GeV\,cm^{-3}s^{-1}}$ and the higher values of $C'(t)= (4.157 - 2.952 t) \cdot 10^{12} \,\rm{GeV\,cm^{-3}s^{-1}}$. All of these values require the same total injection over the source's time scale, an interval of one second. Rigidity-dependent escape was implemented by taking the advective escape as the nominal case (whereby all species propagate one light-second until escape) and two other diffusion cases: Bohm and Kolmogorov. In both cases, the escape is exponentially distributed, $F_{\rm esc}=1 - \exp(-t/t_{\rm diff}(R))$, with diffusion time scales $t_{\rm diff}(R) = 3\cdot10^6 / R$ for the Bohmian case (where the diffusion coefficient is proportional to the rigidity), and $t_{\rm diff}(R) = 2\cdot10^2 / R^{1/3}$ to model the Kolmogorov case. These choices of $F_{\rm esc}(t, R)$ illustrate the changes in the crossing time through the source medium, with a broadening at lower rigidities, as expected for diffusive propagation.

\section{Efficiency over a Monte Carlo approach}
\label{sec:efficiency}

The stochastic framework presented here offers a more efficient method for computing quantities related to UHECR interactions in astrophysical scenarios. Monte Carlo methods, commonly employed for UHECR propagation, have known drawbacks regarding computational efficiency in comparison to the evaluation of closed-form expressions. Nevertheless, for the sake of completeness, we compare the efficiency of this approach to that of a generic Monte Carlo.

Determining the computational cost or efficiency of Monte Carlo methods is not trivial because the conclusion depends on the aspects taken into consideration, e.g., the desired precision, the convergence of the employed algorithm(s), and the specific scenario being simulated. For simplicity, we focus on a few quantities that characterize the efficiency in terms of time and the number of computational operations required. We omit estimating energy costs in this comparison. However, it is clear that, in general, Monte Carlo methods are a poor choice when rare events are of interest and the region of interest in the input phase space is unknown. Additionally, costs are also incurred in training and testing simulations, which are often discarded due to errors. Many of these drawbacks can be addressed with specific sampling techniques or by introducing weights and biases. However, these techniques still require exploratory simulations to understand the underlying phase space.
 
\begin{figure}
     \centering
     \includegraphics[width=.6\textwidth]{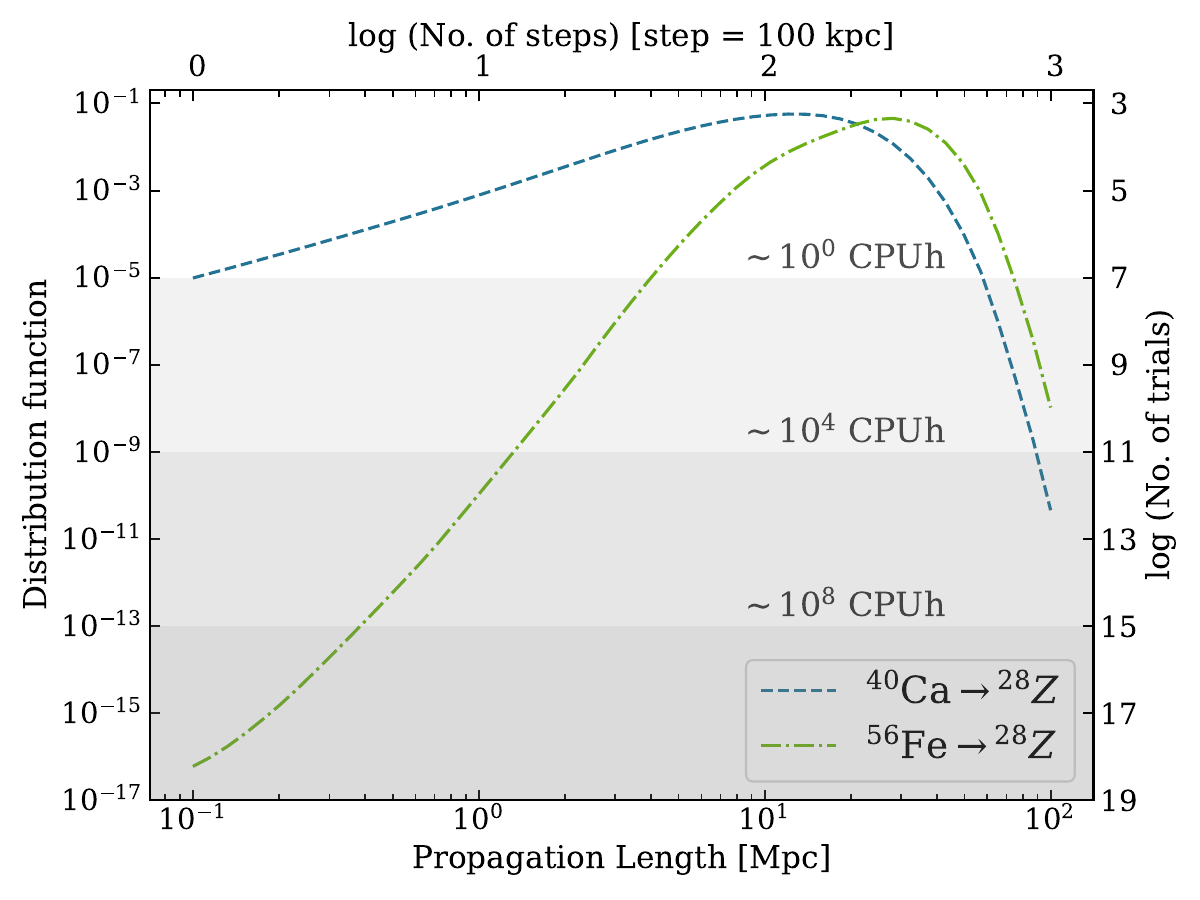}
     \caption{Estimate of the computational effort with a Monte Carlo approach needed to access the probability distributions shown by the lines by evaluating the analytic expressions in this paper (a few seconds for one CPU). The shaded bands correspond to portions of the distributions that can be accessed at the stated computational cost in CPU hours.}
     \label{fig:efficiency_estimation}
 \end{figure}

The goal here is to establish an ideal limit based on two simple assumptions: a) All Monte Carlo trials have a similar computational cost; b) Sampling the input parameter space results in uniform sampling of the desired quantity. These ensure the probability of obtaining values of the distribution in the range $[x_a, x_b]$ is $F(x_b)-F(x_a)$, where $F(x)$ is the distribution function, and for an infinitesimal range $dx$ the probability is given by $f(x)dx$, where $f(x)$ is the density function. 
Figure~\ref{fig:efficiency_estimation} illustrates the computational costs for two different distributions of the probability density of producing nuclei with $A=28$ for $\gamma=5\cdot10^9$: one starting with $^{40}$Ca (dashed blue) and one starting with $^{56}$Fe (dotted-dashed green). The disintegration cross-sections used are from the default table provided in CRPropa~3.2~\citep{Kampert:2012fi,2022JCAP...09..035A} (184 species). The number of trials, given on the right axis, is estimated by considering that, to limit the uncertainty to $\sim10$\%, the number of successful events needed is $N_s\sim100$. This implies the number of trials is $N_t=N_s / p$, where the probability is $p=f(x)dx$, and $f(x)$ the theoretical probability density illustrated with the lines. In terms of CPU time, the computational cost of a Monte Carlo approach can be estimated by assuming $\sim10^{-7}$ CPU hours per trial. This is a conservative estimate based on the reported values for SimProp~v2r4~\citep{Aloisio_2017}. The shaded bands enclose the probability ranges that can be probed with the stated computational effort in CPU hours. A relatively good characterization of both curves can be achieved with a Monte Carlo approach for $\sim1$~CPU hour. However, obtaining the theoretical distributions with the desired 10\% level of uncertainty would require $\gtrsim 10^8$ CPU hours. Nevertheless, these distributions can be computed in a few seconds on a typical laptop using Python scripts and standard mathematical functions with the approach presented in this paper.
 
A different estimate of the computational cost per trial can be made based on the number of propagation steps. In a Monte Carlo propagation of cosmic rays, the distribution over the propagation length is usually obtained through a sequence of propagation steps (of a user defined length), which involves repeatedly testing whether an interaction occurs after each iteration. The upper axis in Fig.~\ref{fig:efficiency_estimation} presents the number of steps required to reach a given propagation length, assuming a step size of 100 kpc. The most likely events involve $\sim100$ steps, a number similar to the average number of steps per trial. Having an estimate of the CPU time for each step allows us to place constraints on the limits of different portions of the distribution and shows that the efficiency depends on whether the step size is much smaller than the length scales of the most relevant probable events. However, the resolution achievable with a given step size is limited since, in a Monte Carlo approach, the step size is generally chosen without knowledge of the form of the distribution. While a smaller step size would yield a better resolution of the entire distribution, it would increase the number of steps simulated with correspondingly more CPU time. Additionally, the asymmetry of these distributions implies that no single step size can describe both the small-scale rise and the large-scale tails with the same precision, without considerable CPU time. In scenarios of interest, it is often not possible to find an optimal step size that describes all underlying distributions with comparable level of precision, as the rise and decrease of different distributions can differ by orders of magnitude and peak at different distances. This occurs not only for different injected species, but also for different values of the Lorentz boost which are commonly simulated with the same steps. These limitations of Monte Carlo methods can be overcome when the underlying distributions are known in closed form.
\subsection{Concrete examples: Comparison to CRPropa}
\begin{figure}
    \centering
    \includegraphics[width=.9\linewidth]{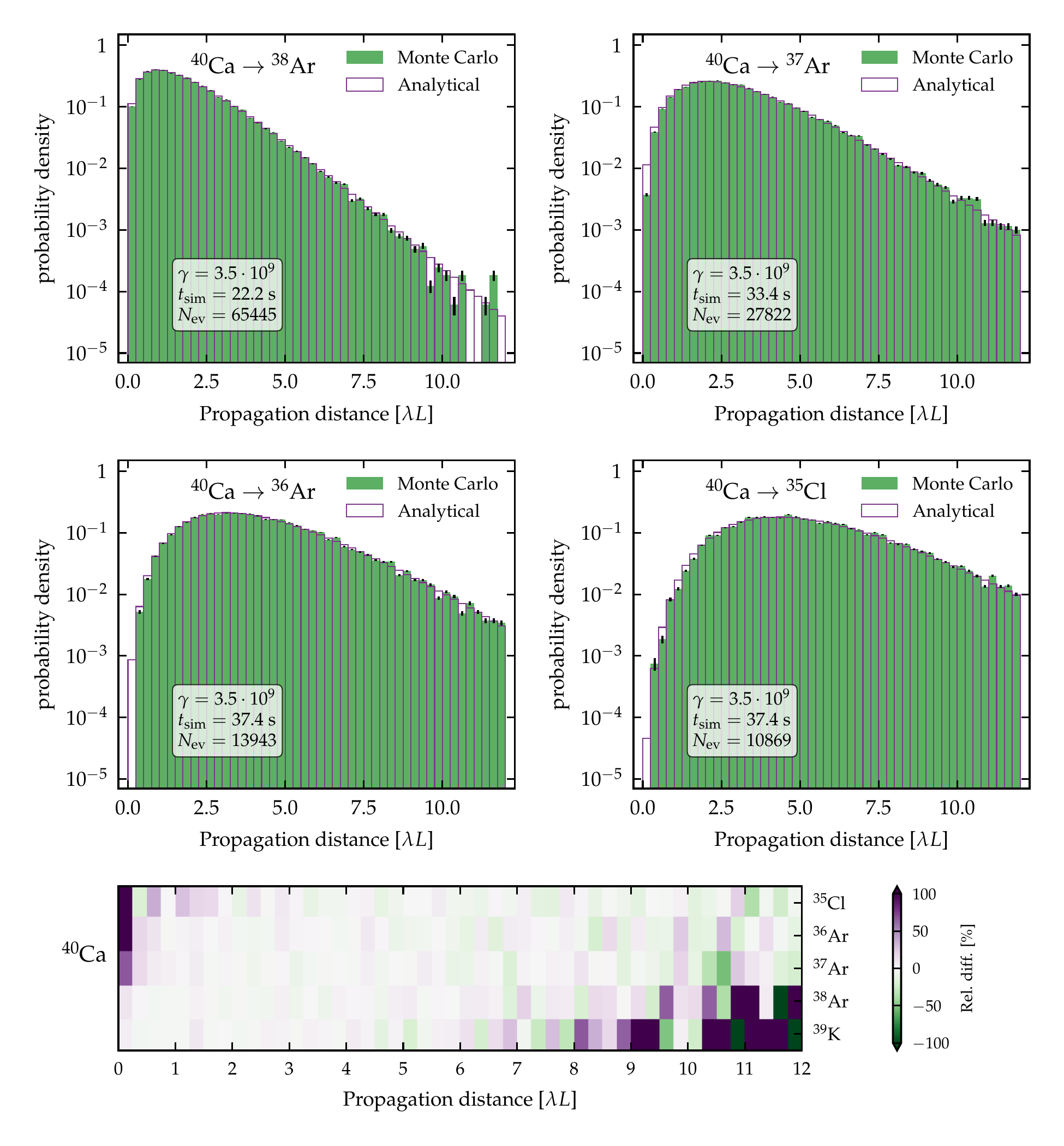}
    \caption{Comparison of CRPropa simulations and analytic distributions for a serial cascade involving nuclear species \{$^{40}$Ca$\to$$^{39}$K$\to$$^{38}$Ar$\to$$^{37}$Ar$\to$$^{36}$Ar$\to$$^{35}$Cl\}. The four top panels show the probability distributions as a function of distance for cases with 2-5 nucleon losses. The bottom panel shows the relative differences between CRPropa and the analytic distributions. There is a row for each case, including the one-nucleon-loss distributions. Each case simulated $10^5$ nuclei with $\gamma=3.5\cdot10^9$. The interaction length for $^{40}$Ca is the reference scale $1/\lambda$.}
    \label{fig:CRISP_CRPropa_comparison}
\end{figure}

Figure~\ref{fig:CRISP_CRPropa_comparison} shows a comparison of some example serial distributions. The simulated nuclear chain is a serial cascade consisting of the following species \{$^{40}$Ca$\to$$^{39}$K$\to$$^{38}$Ar$\to$$^{37}$Ar$\to$$^{36}$Ar$\to$$^{35}$Cl\} and their respective interaction lengths \{8.98, 7.57, 11.48, 9.78, 10.64\} Mpc for a boost of $\gamma=3.5\cdot10^9$. 
The four top panels of the figure show the distributions of the propagation distances needed to disintegrate two, three, four, and five nucleons from the initial $^{40}$Ca nuclei. Results of CRPropa simulations are shown with solid green bars, and results of the analytical approach are shown in hollow purple bars. In the Monte Carlo approach, $10^5$ nuclei were simulated for each case with a fixed propagation step of $\lambda^{-1}/50\approx 180\,$kpc. Here, $\lambda$ is the interaction rate for $^{40}$Ca at the chosen boost $3.5 \cdot 10^9$, which is the shortest for all species in the chain. The expected distance is larger the more nucleons are required to disintegrate, and the distributions become broader. Additionally, the number of recorded nuclei ($N_{\rm ev}$) decreases for distributions with a larger nucleon loss because the number of interactions needed to reach the final state increases (one nucleon is lost per interaction) and the final state is increasingly less probable, consequently increasing the simulation time.
The Monte Carlo simulations closely resemble the analytical distributions within the range of distances with the highest probabilities and are less accurate elsewhere. The bottom panel of Figure~\ref{fig:CRISP_CRPropa_comparison} shows the relative differences between the CRPropa simulations and the analytic distributions. Each row shows the differences for a distribution identified by the final species denoted on the right side, including the one-nucleon-loss case. This demonstrates the good agreement in the range of the highest probabilities, while larger deviations concentrate in the tails of the distributions especially when the probabilities fall below 10$^{-3}$. The simulation times ($t_{\rm sim}$) are considerably longer than the time needed to evaluate the analytic expressions (ranging tens of milliseconds).

\begin{figure}
    \centering
    \includegraphics[width=\linewidth, trim={0 5mm 0 10mm}, clip]{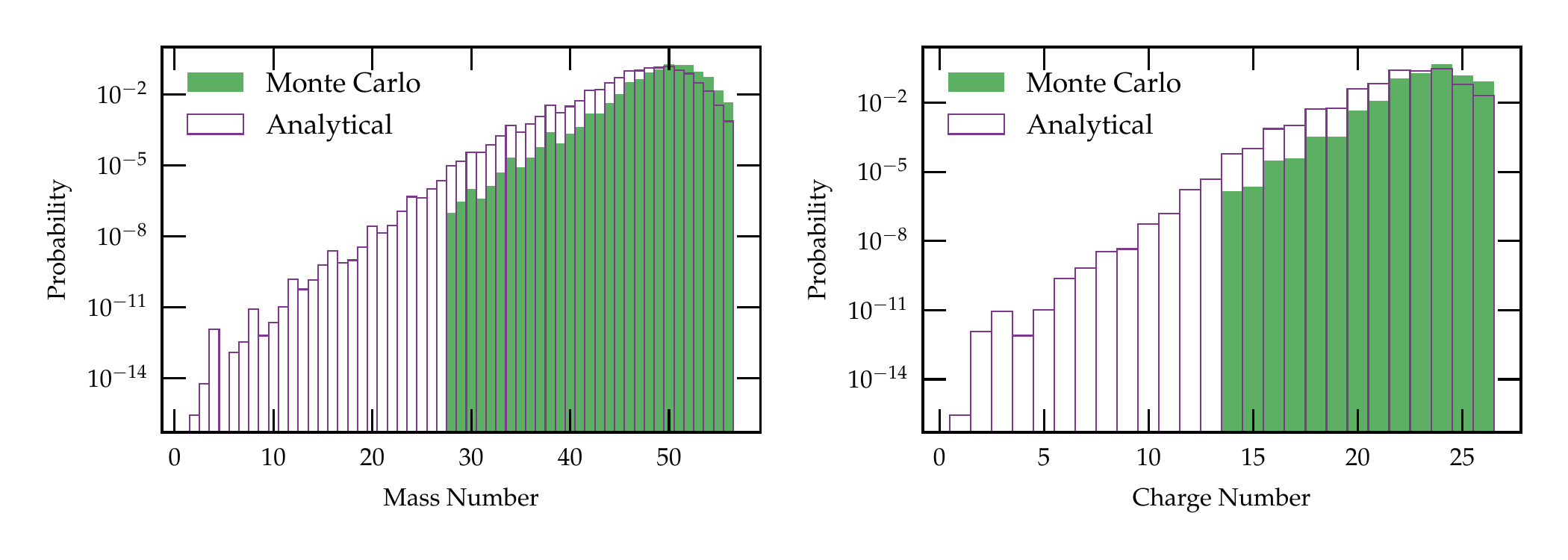}
    
    \includegraphics[width=.7\linewidth, trim={0 10mm 0 20mm}, clip]{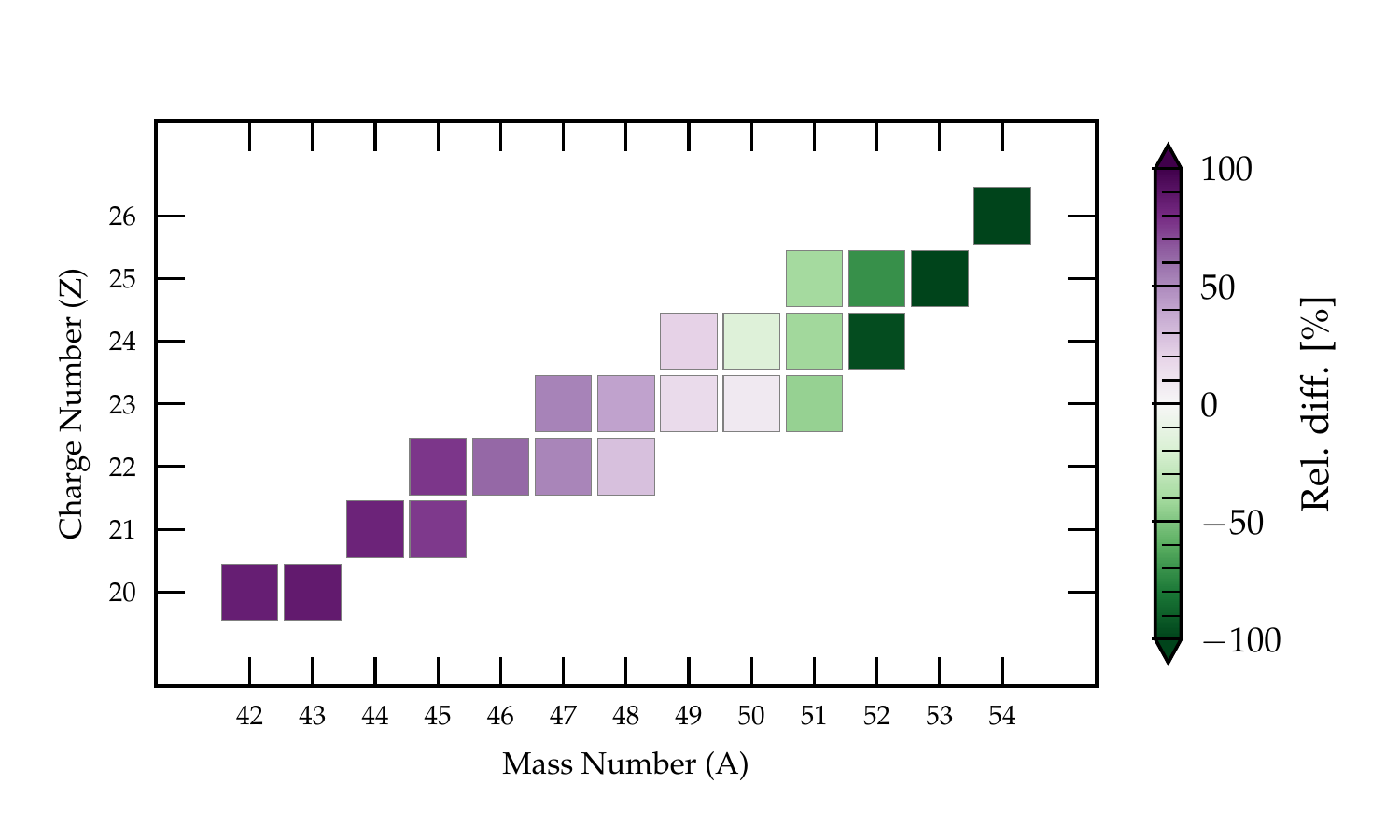}
    \caption{Distribution of heavier secondary species produced in the disintegration of Fe nuclei at boost $\gamma=3.5\cdot 10^9$ after propagating for 30~Mpc. Top: Marginal distributions of mass number (left) and charge number (right) obtained with Monte Carlo (using CRPropa, green solid) versus Analytical (using CRISP, purple lines). Down: Relative differences, in percentages, of Monte Carlo and Analytical probabilities for each secondary species with a probability larger than 1\%.}
    \label{fig:species_distribution}
\end{figure}

Figure~\ref{fig:species_distribution} illustrates concurrent cascades with the distributions of heavier secondary species produced in the disintegration of $^{56}$Fe nuclei after propagating for 30~Mpc with Lorentz boost $\gamma=3.5\cdot 10^9$. 
As in the previous figure, CRPropa simulations are shown with solid green bars and the analytical approach is shown in hollow purple bars. Although the distributions are closer for the most probable bins of mass (charge), the marginal distributions obtained with CRPropa peak at larger masses, indicating a less efficient disintegration. The relative differences between Monte Carlo and the analytical methods are shown in the bottom panel, for secondary species with probability larger than 1\%. 
The most probable species are those with masses around 49-50, and the smallest differences between the distributions are found in that region, while above and below the range the distributions rapidly diverge. The origin of these discrepancies may reflect the choice of a fixed propagation step in CRPropa, which is a user-provided argument that affects the number of interaction samplings performed in propagating a nucleus and its secondaries. A proper choice of step size that is suitable for all intermediate species is not trivial as previously discussed, as each species has different interaction rates (approximately proportional to the mass). Alternatively, allowing CRPropa to make corrections to the step size during simulation can introduce biases in the computed distributions. A detailed study of these discrepancies would require more rigorous comparisons that systematically explore the effects of different parameters and simulation inputs. Such investigations will be pursued in future works.

\end{document}